\documentclass[conference,10pt]{IEEEtran}
\usepackage{amsmath,
            amsfonts,
            amssymb,
            latexsym,
            setspace,
            helvet}
\usepackage{extarrows}
\usepackage{amssymb}
\usepackage{graphicx}
\usepackage{color}
\usepackage{tikz}
\usetikzlibrary{arrows,automata}
\usepackage[latin1]{inputenc}
\usepackage{subfigure}
\usepackage{extarrows}

\usepackage{url}
\urldef{\mailsa}\path|{alfred.hofmann, ursula.barth, ingrid.haas, frank.holzwarth,|
\urldef{\mailsb}\path|anna.kramer, leonie.kunz, christine.reiss, nicole.sator,|
\urldef{\mailsc}\path|erika.siebert-cole, peter.strasser, lncs}@springer.com|

\begin{document}

\newtheorem{theorem}{Theorem}[section]
\newtheorem{lemma}[theorem]{Lemma}
\newtheorem{definition}[theorem]{Definition}
\newtheorem{proposition}[theorem]{Proposition}

\def\squareforqed{\hbox{\rlap{$\sqcap$}$\sqcup$}}
\def\qed{\ifmmode\squareforqed\else{\unskip\nobreak\hfil
\penalty50\hskip1em\null\nobreak\hfil\squareforqed
\parfillskip=0pt\finalhyphendemerits=0\endgraf}\fi}

\newcounter{statement}
\def\stmnum{\hbox to .01pt{}\rlap{\rm \hskip -\displaywidth\thestatement.}}
\def\stm{\refstepcounter{statement}\topsep 2pt \trivlist \item[]\leavevmode
\hbox to\linewidth\bgroup $ \displaystyle \hskip\leftmargini}
\def\endstm{$\hfil \displaywidth\linewidth\stmnum\egroup \endtrivlist}

\newenvironment{proof}{%
\begin{list}{}{\setlength{\topsep}{\jot}\setlength{\parsep}{\topsep}%
\addtolength{\parsep}{-0.3\parsep}\setlength{\leftmargin}{0pt}}%
\parindent 4ex
\item[]\setcounter{statement}{0}\textbf{Proof:}}{\end{list}}

\newcommand{\scarrow}[2]
      { \mathrel{\setbox0 \hbox{ ${\scriptstyle #1}$ }\displaystyle
            \mathop{\hbox to \wd0{$\Longrightarrow$}}\limits_{#1}^{#2}}\,\,\,
      }
\newcommand{\lscarrow}[2]
      { \mathrel{\setbox0 \hbox{ ${\scriptstyle #1}$ }\displaystyle
                 \mathop{\hbox to \wd0{$=\!=\!\Longrightarrow$}}\limits_{#1}^{#2}}
      }
\newcounter{LLN}
\newcommand{\beginit} {
                         \begin{list}{\hfill{\rm \arabic{LLN}.} }{\usecounter{LLN}
                         \setlength{\leftmargin}{\leftmarginii}
                         \setlength{\itemsep}{\smallskipamount}}}

\newbox\mystrutbox
\setbox\mystrutbox=\hbox{\vrule height10pt depth4pt width0pt}
\def\mystrut{\relax\ifmmode\copy\mystrutbox\else\unhcopy\mystrutbox\fi}
\newcommand{\lprf}[3]
       {\llap{{$#3$}\thinspace}{{\mystrut\displaystyle #1}
          \over
         {\mystrut\displaystyle #2}}}
\newcommand{\rprf}[3]
       {{{\mystrut\displaystyle #1}
          \over
         {\mystrut\displaystyle #2}}{\rlap{{$#3$}\thinspace}}}

\newcommand{\dotminus}{-\!\!\!\!^{\textstyle.}\!\!\!\!-}
\newcommand{\lit}[1]{{\rm [#1]}}
\newcommand{\henv}[1]{#1}
\newcommand{\setof}[2]{\{ #1 \: | \: #2 \}}
\newcommand{\bigsetof}[2]{\bigl\{ #1 \: \bigm| \: #2 \bigr\}}
\newcommand{\arrow}[1]{\stackrel{#1}{\longrightarrow}}
\newcommand{\sarrow}[1]{\stackrel{#1}{\Longrightarrow}}
\newcommand{\dobarrow}[1]{\stackrel{#1}{\Longrightarrow}}
\newcommand{\proc}{Pr}
\newcommand{\af}[1]{\forall{\rm F\;}#1}
\newcommand{\ef}[1]{\exists{\rm F\;}#1}
\newcommand{\ag}[1]{\forall{\rm G\;}#1}
\newcommand{\eg}[1]{\exists{\rm G\;}#1}
\newcommand{\varu}[3]{U^{#1}_{#2,#3}}
\newcommand{\varusfg}{\varu{s}{F}{G}}
\newcommand{\varuwfg}{\varu{w}{F}{G}}
\newcommand{\semu}[3]{{\cal U}^{#1}_{#2,#3}}
\newcommand{\semuwfg}{\semu{w}{F}{G}}
\newcommand{\semusfg}{\semu{s}{F}{G}}
\newcommand{\transu}[3]{{\cal T}^{#1}_{#2,#3}}
\newcommand{\transusfg}{\transu{s}{F}{G}}
\newcommand{\transuwfg}{\transu{w}{F}{G}}
\newcommand{\comp}[1]{{\cal C}(#1)}
\newcommand{\prc}[1]{{\cal P}(#1)}
\newcommand{\comppred}[1]{\prec_{#1}}
\newcommand{\bisim}[1]{\sim_{#1}}
\newcommand{\fat}[1]{\mbox{\bf #1}}
\newcommand{\tr}{\mbox{\rm tt}}
\newcommand{\fa}{\mbox{\rm ff}}
\newcommand{\act}{Act}
\newcommand{\impliess}{\: \Rightarrow \:}
\newcommand{\implied}{\Leftarrow}
\newcommand{\implieds}{\:\Leftarrow\:}
\newcommand{\iffs}{\:\Leftrightarrow\:}
\newcommand{\hviss}{\Leftrightarrow}
\newcommand{\sigent}{\sigma_{\entail{}}}
\newcommand{\ua}[2]{#1 \subseteq \sem{\D}#1 \cup #2}
\newcommand{\id}{\mbox{\rm Id}}
\newcommand{\powerset}[1]{{\Large \wp} (#1)}
\newcommand{\sem}[1]{\lbrack\!\lbrack #1 \rbrack\!\rbrack}
\newcommand{\abs}[1]{|\!| #1 |\!|}
\newcommand{\may}[1]{\langle #1 \rangle}
\newcommand{\wmay}[1]{\langle\!\langle #1 \rangle\!\rangle}
\newcommand{\wmust}[1]{\sem{#1}}
\newcommand{\until}[1]{[ #1 \rangle}
\newcommand{\smay}[1]{\langle\!\cdot #1 \cdot\!\rangle}
\newcommand{\must}[1]{[ #1 ]}
\newcommand{\smust}[1]{\lbrack\!\cdot #1 \cdot\!\rbrack}
\newcommand{\sat}[1]{\models_{#1}}
\newcommand{\mx}{{\rm max}}
\newcommand{\mn}{{\rm min}}
\newcommand{\satmn}{\sat{\mn}}
\newcommand{\satmx}{\sat{\mx}}
\newcommand{\sigmax}{\sigma_{\mx}}
\newcommand{\sigmin}{\sigma_{\mn}}
\newcommand{\entail}[1]{\vdash_{#1}}
\newcommand{\entmn}{\entail{\mn}}
\newcommand{\entmx}{\vdash}
\newcommand{\satt}[1]{|\!\!\!\equiv_{#1}}
\newcommand{\sattmx}{\satt{\mx}}
\newcommand{\sattmn}{\satt{\mn}}
\newcommand{\Dtu}{\D^{\tau}}
\newcommand{\Dtd}{\D_{\tau}}
\newcommand{\da}[2]{\semd #1 \cap #2 \subseteq #1}
\newcommand{\ass}[2]{#1 : #2}
\newcommand{\MM}[1]{{\cal M}_{#1}}
\newcommand{\D}{{\cal D}}
\newcommand{\mmid}{{\cal M}_{{\rm Id}}}
\newcommand{\og}{\wedge}
\newcommand{\eller}{\vee}
\newcommand{\semd}{\sem{\D}}
\newcommand{\M}{{\cal M}}
\newcommand{\infrule}[3]
           {\parbox{2cm}{ $$ {\frac {#1}{#2}}\hspace{.5cm}{#3} \hfill $$}}
\newcommand{\infrulegen}[4]
           {{#1}\hspace{.5cm}{\frac {#2}{#3}}\hspace{.5cm}{#4}}
\newcommand{\ent}{\entail{}}
\newcommand{\Gammah}{\widehat{\Gamma}}
\newcommand{\inv}[1]{\textsf{\rm Inv}(#1)}
\newcommand{\pos}[1]{\mbox{\rm Pos}(#1)}
\newcommand{\inva}{\inv{\may{a}\tr}}
\newcommand{\posa}{\pos{\must{a}\fa}}
\newcommand{\op}{{\cal O}}
\newcommand{\even}[1]{\mbox{\rm Even}(#1)}
\newcommand{\unic}{\bigr( \lbr\;|\;\hole\bsl p\bigr) \bsl\coin,\cof}
\newcommand{\live}{\mbox{\rm Live}}
\newcommand{\con}{\mbox{\rm Con}}
\newcommand{\com}{\mbox{\rm Com}}
\newcommand{\scom}{\mbox{{\cal Sc}}}
\newcommand{\dead}{\mbox{\rm Dead}}
\newcommand{\diver}{\mbox{\rm Div}}
\newcommand{\sUntil}[2]{\mbox{\rm Unt}^s(#1,#2)}
\newcommand{\wUntil}[2]{\mbox{\rm Unt}^w(#1,#2)}
\newcommand{\saf}[1]{\mbox{\rm Saf}(#1)}
\newcommand{\uni}{\mbox{\sl Uni}}
\newcommand{\staff}{\mbox{\sl Staff}}
\newcommand{\equip}{\mbox{\sl Equip}}
\newcommand{\acc}{\mbox{acc}}
\newcommand{\del}{\mbox{del}}
\newcommand{\wip}[2]{\mbox{\rm wip}(#1,#2)}
\newcommand{\sop}[2]{\mbox{\rm sop}(#1,#2)}
\newcommand{\cuni}{C_{\mbox{\rm uni}}}
\newcommand{\lbr}{\mbox{\rm lbr}}
\newcommand{\carrow}[2]{\raisebox{0.2ex}{$\,\,\,\,\,{#2\atop #1}
\!\!\!\!\!\!\!\!\arrow{}\,\,
$}}
\newcommand{\ccarrow}[2]{\raisebox{0.2ex}{$\,\,\,\,\,\,{#2\atop #1}
\!\!\!\!\!\!\!\!\!\arrow{}\!\!\!\!\!\!\!\!\arrow{}\,\,
$}}
\newcommand{\dccarrow}[2]{\raisebox{0.2ex}{$\,\,\,\,\,\,{#2\atop #1}
\!\!\!\!\!\!\!\!\!\arrow{}\!\!\!\!\!\!\!\!\arrow{}_\C\,\,
$}}
\newcommand{\cdarrow}[2]{\raisebox{0.2ex}{$\,\,\,\,\,\,{#2\atop #1}
\!\!\!\!\!\!\!\!\!\arrow{}_\C\,
$}}
\newcommand{\dcarrow}[2]{{\raisebox{-1.2ex}{
                         $\stackrel{#2}{\stackrel{\longrightarrow_\Diamond}
                          {\scriptstyle #1}}$  }}}
\newcommand{\rearrow}[2]{\raisebox{-1.2ex}{
                         $\stackrel{#1}{\stackrel{\longrightarrow}
                          {\scriptstyle #2}}$  }}
\newcommand{\bsl}{\backslash}
\newcommand{\hole}{\mbox{$[\;]$}}
\newcommand{\coin}{\mbox{coin}}
\newcommand{\cof}{\mbox{cof}}
\newcommand{\unicp}{\bigr( \cof .(\lbr + p.\lbr) \;|\;\hole\bsl p\bigr)
                                   \bsl \coin,\cof}
\newcommand{\unicpp}{\bigr( (\lbr + p.\lbr)\;|\;\hole\bsl p\bigr)
                             \bsl \coin,\cof}
\newcommand{\synen}[2]{#1\,\,{\bf{\sf with}}\,\,#2}
\newcommand{\emax}[2]{\mbox{{\bf {\sf max }}}#1\,\,{\bf{\sf with}}\,\,#2}
\newcommand{\emin}[2]{\mbox{{\bf {\sf min }}}#1\,\,{\bf{\sf with}}\,\,#2}
\newcommand{\lmax}{\mbox{{\sf max}\,}}
\newcommand{\lmin}{\mbox{{\sf min}\,}}
\newcommand{\Dsem}[1]{{\sf D} \lbrack\!\lbrack #1 \rbrack\!\rbrack}
\newcommand{\Delsem}[1]{{\sf \Delta} \lbrack\!\lbrack #1 \rbrack\!\rbrack}
\newcommand{\Asem}[1]{{\sf E} \lbrack\!\lbrack #1 \rbrack\!\rbrack}
\newcommand{\Lsem}[1]{{\sf L} \lbrack\!\lbrack #1 \rbrack\!\rbrack}
\newcommand{\Nsem}[1]{{\sf N} \lbrack\!\lbrack #1 \rbrack\!\rbrack}
\newcommand{\DMTSt}{\mbox{DMTS$^2$}\,}
\newcommand{\DMTSs}{\mbox{DMTS$^\ast$}\,}
\newcommand{\conf}[2]{\langle #1,#2\rangle}
\newcommand{\chop}[1]{\mbox{{\tt chop}$(#1)$}}
\newcommand{\tl}[1]{\mbox{{\tt tl}$(#1)$}}
\newcommand{\eemax}[1]{\mbox{{\bf {\sf max }}}#1}
\newcommand{\eemin}[1]{\mbox{{\bf {\sf min }}}#1}
\newcommand{\pre}{\mbox{\,{\tt pre}\,}}
\newcommand{\conft}[1]{\langle #1\rangle}
\newcommand{\incon}{IC}
\newcommand{\hml}{{\cal M}}
\newcommand{\depth}{depth}
\newcommand{\dsat}{{\,\mbox{{\tt sat}}}}
\newcommand{\branb}{{\approx_b}}
\newcommand{\V}{{\cal V}}
\newcommand{\Z}{{\cal Z}}
\newcommand{\ins}[3]{\vdash^?_{#3}#1\colon#2}
\newcommand{\insp}[3]{\vdash^{#3}#1\colon#2}
\newcommand{\insn}[3]{\not\vdash_{#3}#1\colon#2}
\newcommand{\wi}[2]{{\cal W}(#1,#2)}


\newcommand{\DEF}               {\stackrel{\rm def}{=}}
\newcommand{\invo}              {\textsf{inv}}
\newcommand{\res}               {\mathit{res}}
\newcommand{\lin}               {\textsf{lin}}
\newcommand{\Abs}               {\textsf{Abs}}
\newcommand{\client}            {\textsf{Client}}
\newcommand{\Sys}               {\textsf{Sys}}
\newcommand{\Systau}            {\textsf{Sys1}}
\newcommand{\Sysinv}            {\textsf{Sys2}}
\newcommand{\Sysres}            {\textsf{Sys3}}
\newcommand{\Sysobj}            {\textsf{Sys4}}
\newcommand{\emp}               {\varepsilon}
\newcommand{\Ex}                {\bf Ex}
\newcommand{\Eva}               {\bf Eva}
\newcommand{\true}              {\mathit{true}}
\newcommand{\false}             {\mathit{false}}
\newcommand{\IF}                {\textsf{if}}
\newcommand{\ELSE}              {\textsf{else}}
\newcommand{\WHILE}             {\textsf{while}}
\newcommand{\SKIP}              {\textsf{skip}}

\newcommand{\StackOp}           {\mathit{StackOp}}
\newcommand{\LesOp}             {\mathit{LesOp}}
\newcommand{\TryStackOp}        {\mathit{TryStackOp}}
\newcommand{\TryCollision}      {\mathit{TryCollision}}
\newcommand\ignore[1]{}
\newcommand{\s}                 {\textsf{S}}
\newcommand{\dbs}               {\approx_{\bf d}}
\newcommand{\ds}                {\leq_{\bf d}}

\newcommand{\necc}[2]{#1_{\Box}.#2}
\newcommand{\perm}[2]{#1_{\Diamond}.#2}
\newcommand{\rec}{\mbox{\bf rec}}
\newcommand{\barrow}[1]{\stackrel{#1}
                   {\longrightarrow_\Box}}
\newcommand{\cutarrow}[1]{\stackrel{#1}
                   {\longrightarrow_{\odot}}}
\newcommand{\zarrow}[1]{\stackrel{#1}
                   {\Longrightarrow_\Diamond}}
\newcommand{\marrow}[1]{\stackrel{#1}
                   {\longrightarrow_m}}
\newcommand{\darrow}[1]{\stackrel{#1}
                   {\longrightarrow_\Diamond}}
\newcommand{\tilarrow}[1]{
		   {\,\,\,\,\,\,{#1\atop{}}\!\!\!\!\!\!\!\raisebox{-0.2ex}
		{$\sim$}\!\!\!\rightarrow_\Diamond}\,}
\newcommand{\refin}{\lhd}
\newcommand{\rrefin}{\rhd}
\newcommand{\U}{{\cal U}}
\newcommand{\one}{\mbox{\bf 1}}
\newcommand{\nil}{nil}
\newcommand{\R}{{\cal R}}
\newcommand{\F}{{\cal F}}
\newcommand{\W}{{\cal W}}
\newcommand{\WC}{{\cal W}_c}
\newcommand{\E}{{\cal E}}
\newcommand{\semnil}{\mbox{\rm nil}}
\newcommand{\semperm}[1]{\lceil #1_{\Diamond}\rceil}
\newcommand{\semnecc}[1]{\lceil #1_{\Box}\rceil}
\newcommand{\altsem}[1]{\{\!\lbrack #1 \rbrack\!\}}
\newcommand{\finsyn}{{\cal FS}}
\newcommand{\ol}[1]{\overline{#1}}
\newcommand{\C}{{\cal C}}

\newcommand{\csystem}
         {(\langle C_n^m\rangle_{n,m} , A , 
           \langle \arrow{}_{n,m} \rangle_{n,m})}
\newcommand{\proj}[2]{\Pi_{#1}^{#2}}
\newcommand{\ident}[1]{\mbox{I}_{#1}}
\newcommand{\pref}[1]{{#1}.(\,\,)}
\newcommand{\zero}{{\bf 0}}
\renewcommand{\comp}{\mbox{{\rm C}}}
\renewcommand{\nil}{{\bf 0}}
\newcommand{\fdb}[1]{{#1}^{\dagger}}
\renewcommand{\F}{\mbox{{\rm F}}}
\renewcommand{\R}{{\cal R}}
\newcommand{\asscomp}{\mbox{{\rm A}}_{\circ}}
\newcommand{\assprod}{\mbox{{\rm A}}_{\times}}
\newcommand{\distprod}{\mbox{{\rm D}}_{\circ}^{\times}}
\newcommand{\distfdb}{\mbox{{\rm D}}_{\times}^{\dagger}}
\newcommand{\idaxiom}{\mbox{{\rm I}}}
\newcommand{\zerocomp}{\mbox{{\rm Z}}_{\circ}}
\newcommand{\zeroprod}{\mbox{{\rm Z}}_{\times}}
\newcommand{\zerofdb}{\mbox{{\rm Z}}_{\dagger}}
\newcommand{\zerozero}{\mbox{{\rm Z}}\mbox{{\rm Z}}}
\newcommand{\zeronull}{\mbox{{\rm Z}}_0}
\newcommand{\projdef}{\mbox{{\rm P}}_d}
\newcommand{\projprod}{\mbox{{\rm P}}_{\times}}
\newcommand{\fixaxiom}{\mbox{{\rm F}}}
\renewcommand{\zero}[2]{{\cal O}_{#1}^{#2}}
\renewcommand{\arraystretch}{1.3}
\newcommand{\letmax}[2]{\mbox{{\bf {\sf letmax }}}#1\,\,\mbox{{\bf
{\sf in }}}#2}
\newcommand{\letmix}[2]{\mbox{{\bf {\sf let }}}#1\,\,\mbox{{\bf
{\sf in }}}#2}
\newcommand{\letmin}[2]{\mbox{{\bf {\sf letmin }}}#1\,\,\mbox{{\bf
{\sf in }}}#2}
\newcommand{\form}{{\cal F}}
\newcommand{\decl}{{\cal D}}
\newcommand{\alt}{\,\,\,|\,\,\,}
\newcommand{\Fsem}[1]{{\sf
F} \lbrack\!\lbrack #1 \rbrack\!\rbrack}
\newcommand{\Dmax}[1]{{\sf
D} \lbrack\!\lbrack #1 \rbrack\!\rbrack}
\newcommand{\Dmin}[1]{{\sf D} \lbrack\!\lbrack #1 \rbrack\!\rbrack}
\renewcommand{\tr}{\mbox{{\sf tt}}}
\renewcommand{\fa}{\mbox{{\sf ff}}}
\newcommand{\func}{\rightarrow}
\newcommand{\env}{{\cal E}}
\renewcommand{\sat}{\models}
\newcommand{\w}{{\cal W}}
\newcommand{\vc}{V^{\C}}
\newcommand{\valid}{\models}
\newcommand{\wvalid}{\models_{\times}}
\newcommand{\LC}{{\cal L}}
\newcommand{\RC}{{\cal R}}
\newcommand{\uu}[2]{\mbox{$\bigcup_{#1} #2$}}
\newcommand{\nn}[2]{\mbox{$\bigcap_{#1} #2$}}
\newcommand{\fuu}[2]{\mbox{$\bigvee_{#1}#2$}}
\newcommand{\fnn}[2]{\mbox{$\bigwedge_{#1}#2$}}
\newcommand{\Dmaxsem}[2]{{\sf D}_\nu\sem{#1}#2}
\newcommand{\Dminsem}[2]{{\sf D}_\mu\sem{#1}#2}
\newcommand{\B}{{\cal B}}
\newcommand{\maxd}{{\sf max}}
\newcommand{\mind}{{\sf min}}

\renewcommand{\powerset}[1]{{\sf 2}^{#1}}
\renewcommand{\S}{{\cal S}}
\newcommand{\rest}[1]{\backslash #1}
\newcommand{\T}{{\cal T}}
\newcommand{\ttb}{\preceq}
\newcommand{\nestbis}[1]{\simeq_{#1}}
\newcommand{\N}{{\cal N}}
\renewcommand{\P}{{\cal P}}
\newcommand{\weight}{{w}}
\newcommand{\ifthel}[3]{{\mbox{\sf if} \,\,#1\,\,\mbox{\sf then}\,\,#2\,\,\mbox{\sf else}\,\,#3}}
\newcommand{\grdsim}{\stackrel{.}{\sim}}
\newcommand{\fn}{{\tt fn}}
\newcommand{\fresh}{{\tt fresh}}

%


\title{A Complete Axiomatisation for \\Divergence Preserving Branching Congruence
of Finite-State Behaviours
}



\author{\IEEEauthorblockN{Xinxin Liu}
\IEEEauthorblockA{State key laboratory of Computer Science, ISCAS\\
University of Chinese Academy of Sciences\\
RISE, School of Computer and Information Science\\
South West University, China\\
Email: xinxin@ios.ac.cn}
\and
\IEEEauthorblockN{Tingting Yu}
\IEEEauthorblockA{Beijing Sunwise information Technology Ltd\\
Beijing Institute of Control Engineering\\
Email: yutingting@sunwiseinfo.com}}

\IEEEoverridecommandlockouts
\IEEEpubid{\makebox[\columnwidth]{978-1-6654-4895-6/21/\$31.00~
\copyright2021 IEEE \hfill} \hspace{\columnsep}\makebox[\columnwidth]{ }}

\maketitle

\begin{abstract}
We present
an equational inference system  for finite-state expressions, and prove that the system
is sound and complete with respect to divergence preserving branching congruence, closing a problem that
has been open since 1993. The inference system refines Rob van Glabbeek's simple and elegant
complete axiomatisation for branching bisimulation congruence of finite-state behaviours
by joining four simple axioms
after dropping one
axiom which is unsound under the more refined divergence sensitive semantics.
\end{abstract}

\IEEEpeerreviewmaketitle

\section{Introduction}

Over the years the notion of bisimulation which was proposed by Park
and popularized by the work of Milner emerges as a very important
foundation for concurrency theory. Based on this notion, many
interesting equivalence and congruence relations are introduced and studied.
Rob van Glabbeek gave a fairly complete list of these equivalences in
\cite{glab90a} and
\cite{glab90b}.

For a bisimulation based congruence relation
on a set of expressions,
an interesting question is whether there is an equational inference
system, or axiomatisation, which infers exactly the pairs of
equal expressions.
Even in the  cases where the equivalence relation
is decidable, an inference system of this nature is still important
since it conveys valuable
information about the rationale behind the equalities. For finite-state
expressions (expressions which can only generate finitely many states
but may generate infinite behaviours),
Milner pioneered this line of research, and proposed complete
axiomatisations for strong bisimulation congruence
\cite{milner84} and observational bisimulation congruence
\cite{milner89}. Following Milner's work, Walker \cite{walker90} and
Lohrey et al. \cite{lohrey05} proposed
complete axiomatisations for variations of observational bisimulation congruence
which take divergence
behaviour into account, and
van Glabbeek
proposed a complete axiomatisation for branching bisimulation congruence \cite{glabb96}.

The notion of branching bisimulation was introduced by van Glabbeek and Weijland
in \cite{vonglaweij96}, in which a refined notion of divergence preservation is
introduced and used to define divergence-preserving version of
the corresponding bisimulation equivalence and congruence. So, in \cite{glabb96}
van Glabbeek posed the following natural question: to find
a complete axiomatisation for the divergence-preserving version of branching
bisimulation congruence. The problem remains open until today, although
Chen and Lu \cite{chen08} and Fu \cite{fu2015} proposed complete axiomatisations for
divergence-preserving semantics for sub-languages of finite-state behaviours. In this paper,
we propose an axiomatisation for finite-state behaviours, and prove its soundness and
completeness with respect to divergence-preserving branching bisimulation congruence.

The divergence-preserving branching bisimulation equivalence with the corresponding
congruence is unique in that
it is the finest possible bisimulation equivalence which abstracts from internal moves.
Thus, a complete axiomatisation for the congruence could serve
as a core theory which can be readily extended to axiomatisations for other
bisimulation based congruences by adding new axioms.

For the completeness proof, we use Milner's \cite{milner84} framework of set of guarded equations, with
the following difference: instead of using the product construction of equation sets proposed by Milner,
we use the quotient construction which was introduced by Grabmayer and
Fokkink in \cite{GF20} and independently by  Liu and Yu in \cite{LY20}.

The paper is organized as follows. In the next section we settle the preliminaries,
including the definitions and properties of the equivalence and congruence relations.
In section III we present the inference system and prove its soundness for divergence-preserving
branching congruence. In section IV we introduce the notion of standard sum, and
prove a standardization result:
every expression can be proven equal to a standard sum.
In section V we study standard equation systems (SES), and prove the quotient theorem, i.e.
equivalent formal variables of an SES have common provable solution in a related guarded
equation system.
Using the result of section V,
the completeness of the inference system is proved in section VI.
Then, we
conclude in section VII.

\section{Expressions, Divergence-Preserving Semantics}

Let ${\cal V}$ be an infinite set of {\em variables},
${\cal A}$ be an infinite set of  {\em visible actions},
$\tau$ be the {\em invisible action} or {\em silent move} ($\tau\not\in {\cal A}$).
We write ${\cal A}_\tau$ for ${\cal A}\cup\{\tau\}$. Consider the
set $\E$ of {\em process expressions}, given by the following BNF rules:
$$\begin{array}{ccc|c|c|c|c}
E&::=& {\bf 0} &X&a.E& E+E&\mu X.E
\end{array}
$$
where $a\in{\cal A}_\tau$, $X\in {\cal V}$. The precise meaning of the
expressions will be given by operational semantics later.
Here we provide the following
explanations for the syntax, which may help to
understand the intuitive meaning of the expressions:
\begin{itemize}
\item
${\bf 0}$ is the expression which is not capable of any
action;
\item $a.E$ is a prefix expression which first performs the action $a$ and
then proceeds as $E$;
\item
$E+F$ is a non-deterministic expression which is capable of actions from $E$ and $F$;
\item $\mu X.E$ is a recursion which behaves as
$E$ except that whenever $X$ is encountered in an execution then the rest behaviour
is as $\mu X.E$.
\end{itemize}

We assume the usual notion of free and bound occurrence of
variables with respect to the variable binder $\mu$, write $FV(E)$ for the set of free
variables of $E$,
and write $E\{F/X\}$ for the resulting expression obtained by (capture free) substitution of
$F$ for (free occurrences of) $X$ in $E$. For a set of variables $\{X_1,\ldots,X_n\}$, we write
$E\{F_1/X_1,\ldots,F_n/X_n\}$ for the simultaneous
(capture free) substitution of
$F_1$ for $X_1,\ldots$, $F_n$ for $X_n$ in $E$.
Sometimes we will also use set notation
to write simultaneous
substitution of $F_i$ for $X_i$ in $E$ for each $i\in I$
as $E\setof{F_i/X_i}{i\in I}$, where $I$ is an index set. $\setof{F_i/X_i}{i\in I}$ can also be
used standing alone to represent the intended substitution.
Note that $E\{E_1/X_1\}\{E_2/X_2\}$ stands for the expression obtained by successive substitution of
$E_1$ for the free occurrences of $X_1$ in $E$, and then $E_2$ for the free occurrences of
$X_2$ in $E\{E_1/X_1\}$. We write
$E\equiv F$ when $E,F$ are syntactically identical expressions.

The operational semantics of expressions is given by a transition relation
$\arrow{}$ and a binary relation $\vartriangleright$ between expressions and
variables defined as follows.
\begin{definition}\label{defoftrans}{\sl
The transition relation $\arrow{}\subseteq\E\times {\cal A}_\tau\times\E$
is the smallest relation such that
(we write $E\arrow{a}E'$ for $(E,a,E')\in\arrow{}$):
\begin{enumerate}
\item $a.E\arrow{a}E$;
\item If $E_1\arrow{a}E'$ then $E_1+E_2\arrow{a}E'$;
\item If $E_2\arrow{a}E'$ then $E_1+E_2\arrow{a}E'$;
\item If $E\{\mu X.E/X\}\arrow{a}E'$ then $\mu X.E\arrow{a}E'$.
\end{enumerate}

The relation $\vartriangleright\subseteq\E\times\V$
is the smallest relation such that
(we write $E\vartriangleright X$ for $(E,X)\in\,\vartriangleright$):
\begin{enumerate}
\item $X\vartriangleright X$;
\item If $E_1\vartriangleright X$ then $E_1+E_2\vartriangleright X$;
\item If $E_2\vartriangleright X$ then $E_1+E_2\vartriangleright X$;
\item If $E\{\mu Y.E/X\}\vartriangleright X$ then $\mu Y.E\vartriangleright X$.
\end{enumerate}

Also we write $\dobarrow{}$ for $(\arrow{\tau})^*$
and $\tilde\triangleright$ for $\dobarrow{}\vartriangleright$
(in this paper we write $R^*$ for the reflexive and transitive closure of
a binary relation $R\subseteq\E\times\E$, and write
$R_1R_2$ for the composition of binary relations $R_1$ and $R_2$ when $R_1$'s codomain and
$R_2$'s domain are both $\E$).

In an expression $E$,
an occurrence of a variable $X$ is said {\em guarded} if the occurrence is within
a subexpression $a.E'$ of $E$ where $a\not=\tau$. $X$ is said guarded in $E$
if every free occurrence of $X$ in $E$ is a guarded occurrence.
}
\end{definition}
%

To see some examples, $\tau.X+a.{\bf 0}\ \tilde\triangleright X$,
and if $X\notin FV(E)$ and $a\not=\tau$, then
$X$ is guarded in $\tau.(E+a.(F+\tau.X))$.

For $E\in\E$, an infinite $\tau$-run from $E$ is an infinite sequence of expressions $E_0E_1\ldots E_i\ldots$
such that $E_0\equiv E$ and $E_{i-1}\arrow{\tau}E_i$ for $i>0$.

Next we state some lemmas which are needed later.



\begin{lemma}\label{glabblemma1}
Let $E,F,H\in\E, a\in{\cal A}_\tau, X\in{\cal V}$. Then:
\begin{enumerate}
\item If $H\{E/X\}\arrow{a}F$, then either there is $H'$ such that $H\!\arrow{a}\!H'$ and $F\equiv H'\{E/X\}$,
or $H\!\vartriangleright\!X$ and $E\arrow{a}F$;
\item If $H\{E/X\}\vartriangleright Y$, then either $H\vartriangleright Y$, or $H\vartriangleright X$ and $E\vartriangleright Y$;
\item If $H\arrow{a}H'$, then $H\{E/X\}\arrow{a}H'\{E/X\}$;
\item If $H\vartriangleright X$ and $E\arrow{a}F$ then $H\{E/X\}\arrow{a}F$.
\end{enumerate}
\end{lemma}
The proof of this lemma can be found in \cite{glabb96} (Lemma 4).

\begin{lemma}\label{glabblemma2} Let $X$ be a variable, $H\in\E$. Then $X$ occurs unguarded in $H$
if and only if
$H\,\tilde\triangleright X$.
\end{lemma}
{\bf Proof}
It is easy to prove by induction on the rules defining $\arrow{}$ and $\vartriangleright$
the following:
\begin{enumerate}
\item if $H\vartriangleright X$ then $X$ occurs unguarded in $H$;
\item  if $H\arrow{\tau}H'$ and $X$ occurs unguarded in $H'$ then
$X$ occurs unguarded in $H$.
\end{enumerate}
Note that $H\,\tilde\triangleright X$ is $H\dobarrow{}\vartriangleright\!X$.
Then the "if" direction can be proved by an easy induction on the length of transition sequence
$\dobarrow{}$, and the "only if" direction can be proved
by using Lemma \ref{glabblemma1} and analyzing the structure of $H$.
\hfill\qed


\begin{lemma}\label{glabblemma3} Let $E,F\in\E$, $a\in{\cal A}_\tau$, $X,W\in\V$.
\begin{enumerate}
\item
$\mu X.E\arrow{a}F$ if and only if there is $E'\in\E$ such that $E\arrow{a}E'$ and $F\equiv E'\{\mu X.E/X\}$.
\item $\mu X.E\vartriangleright W$ if and only if $E\vartriangleright W$ and $W,X$ are different variables.
\end{enumerate}
\end{lemma}
{\bf Proof.} The proof of 1) can be found in \cite{lohrey05} (Lemma 6).
The proof of 2) is straightforward.
\hfill\qed

To present the definition of divergence-preserving branching bisimulation, we first define
a number of functions on binary relations, and study the relationships of these functions.

\begin{definition}\label{DefofStrongBisim}{\sl For a binary relation ${R}\subseteq\E\times\E$, define
binary relations ${\cal S}(R), {\cal B}(R), {\cal B}'(R),$ and  ${\cal B}^\vartriangle(R)$ as follows:
\begin{enumerate}
\item ${\cal S}(R)$ is a binary relation such that $(E,F)\in{\cal S}(R)$ if:
\begin{enumerate}
\item whenever $E\!\arrow{a}\!E'$, then there exists $F'$ such that
$F\arrow{a}F'$ and $(E',F')\in R$;
\item whenever $E\vartriangleright\!X$, then  $F\vartriangleright\!X$.
\end{enumerate}
\item ${\cal B}(R)$ is a binary relation such that $(E,F)\in{\cal B}(R)$ if:
\begin{enumerate}
\item whenever $E\!\arrow{a}\!E'$, then either $a\!=\!\tau$ and there
exists
$F'$ s.t.
$F\dobarrow{}F'$and $(E,F'),(E'\!,F')\!\in\!R$, \\
or
there exist $F',F''$ s.t. $F\dobarrow{}F', F'\arrow{a}F''$
and $(E,F'),(E',F'')\in R$;
\item whenever $E\!\vartriangleright\!\!X$, then there exists $F'$such that $F\dobarrow{}F'$, $(E,F')\in R$,
and $F'\vartriangleright\!X$.
\end{enumerate}
\item ${\cal B}'(R)$ is a binary relation such that $(E,F)\in{\cal B}'(R)$ if:
\begin{enumerate}
\item whenever $E\!\arrow{a}\!E'$, then either $a\!=\!\tau$ and there
exists
$F'$s.t.
$F\!\arrow{\tau}\dobarrow{}\!F'$and $(E,F'),(E'\!,F')\!\in\!R$, \\
or
there exist $F',F''$ s.t. $F\dobarrow{}F', F'\arrow{a}F''$
and $(E,F'),(E',F'')\in R$;
\item whenever $E\!\vartriangleright\!\!X$, then there exists $F'$such that $F\dobarrow{}F'$, $(E,F')\in R$,
and $F'\vartriangleright\!X$.
\end{enumerate}
\item ${\cal B}^\vartriangle(R)$ is a binary relation such that $(E,F)\in{\cal B}^{\vartriangle}(R)$ if:
\begin{enumerate}
\item $(E,F)\in{\cal B}(R)$;
\item
whenever $EE_1\ldots E_i\ldots$ is an infinite $\tau$-run from $E$, 
then there exist $E_j$ on the $\tau$-run and $F'$ such that $F\arrow{\tau}\dobarrow{}F'$ and $(E_j,F')\in R$.
\end{enumerate}
\end{enumerate}
}
\end{definition}
${\cal S}, {\cal B}, {\cal B}^\vartriangle$ will be used to define corresponding
bisimulation equivalences.
${\cal B}'$ will be used in introducing a divergence preserving bisimulation verification
technique. Note that ${\cal B}'$ differs from
${\cal B}$ only in one place where ${\cal B'}$ requires $\arrow{\tau}\dobarrow{}$
(in which at least one $\arrow{\tau}$ step must be made)
instead of the transition sequence $\dobarrow{}$. The result of this subtle change
made a big difference as ${\cal B}'(R)$ becomes a subset of not only ${\cal B}(R)$ but also
${\cal B}^\vartriangle(R)$ for any binary relation $R$, as shown by the following lemma.

\begin{lemma}\label{distinguishingpower} Let $R\subseteq\E\times\E$. Then
$${\cal S}(R)\subseteq{\cal B}'(R)\subseteq{\cal B}^\vartriangle(R)\subseteq{\cal B}(R).$$
\end{lemma}
{\bf Proof.} ${\cal S}(R)\subseteq{\cal B}'(R)$ and ${\cal B}^\vartriangle(R)\subseteq{\cal B}(R)$
can be checked immediately from the definitions. To see ${\cal B}'(R)\subseteq{\cal B}^\vartriangle(R)$,
note that from the definitions one easily observes ${\cal B}'(R)\subseteq{\cal B}(R)$. Then
for all $(E,F)\in{\cal B}'(R)$, whenever $EE_1\ldots E_i\ldots$ is an infinite $\tau$-run from $E$,
the transition $E\arrow{\tau}E_1$ would find some $F'$
such that $F\arrow{\tau}\dobarrow{}F'$ and $(E_1,F')\in R$, meeting the requirement of being
a member of ${\cal B}^\vartriangle(R)$.
\hfill\qed

\begin{definition}\label{bbwed}{\sl
A binary relation $R\subseteq\E\times\E$ is a
{\em strong bisimulation} if $R$ is symmetric and $R\subseteq{\cal S}(R)$.

A binary relation $R\subseteq\E\times\E$ is a
{\em branching bisimulation} if $R$ is symmetric and $R\subseteq{\cal B}(R)$.

A binary relation $R\subseteq\E\times\E$ is a
{\em progressing branching bisimulation} if $R$ is symmetric and $R\subseteq{\cal B}'(R)$.

A binary relation $R\subseteq\E\times\E$ is a
{\em divergence-preserving branching bisimulation} if $R$ is symmetric and  $R\subseteq{\cal B}^\vartriangle(R)$.

Define three binary relations, called {\em strong bisimilarity, branching bisimilarity,
divergence-preserving branching bisimilarity}, and
written $\sim$, $\approx_b$, $\approx^\vartriangle_b$ respectively,
as follows
$$\begin{array}{l}
\,\sim\ \,=\bigcup\setof{R}{R\mbox{ is a strong bisimulation }},\\
\approx_b\,=\bigcup\setof{R}{R\mbox{ is a branching bisimulation}},\\
\approx^\vartriangle_b=\bigcup\setof{R}{R\mbox{ is a div.-pre. branching bisimulatoin}}.
\end{array}
$$
}
\end{definition}

For $\sim,\approx_b,\approx^\vartriangle_b$, we have the following justification. 
\begin{theorem}\label{theorem2.3}
$\sim,\approx_b$, and $\approx^\vartriangle_b$ are equivalence relations. Moreover
\begin{enumerate}
\item $\sim$ is the coarsest strong bisimulation;
\item $\approx_b$ is the coarsest branching bisimulation;
\item $\approx^\vartriangle_b$ is
the coarsest divergence-preserving branching bisimulation.
Moreover, it is  the coarsest equivalence relation
which is a branching bisimulation and which preserves divergence, i.e.
for two equivalent expressions if one has an infinite $\tau$-run within
its equivalence class then so is the other.
\end{enumerate}
\end{theorem}
{\bf Proof.} 1), 2) are well known (\cite{milner}, \cite{vonglaweij96}). See \cite{LYZ17} for 3).
\hfill\qed

Note that $\approx_b$ does not respect divergence. A simple example to show this is
$\mu X.(\tau.X+a.{\bf 0})\approx_b\tau.a.{\bf 0}$, where the left expression has an infinite
$\tau$-run, while the right expression has not. Thus $\approx^\vartriangle_b$ provides an alternative
when divergence needs to be taken into account.
There are different presentations of divergence-preserving
branching bisimilarity. It was called
branching bisimilarity with explicit divergence in \cite{vonglaweij96} and \cite{GLT09},
and called complete branching bisimilarity in \cite{LYZ17}. When first introduced in
\cite{vonglaweij96}, it was defined
as the coarsest equivalence relation
which is a branching bisimulation and which preserves divergence. 3) of Theorem \ref{theorem2.3}
shows that the present definition gives
the same relation.
In \cite{GLT09}, a condition similar to condition b) in 4) of Definition \ref{DefofStrongBisim} for matching a divergent run requires that
$F'$ is found after exactly one step of
$\tau$ action. The condition b) in 4) of Definition \ref{DefofStrongBisim}
is from \cite{LYZ17}, which allows $F'$ to be found after one or more steps of
$\tau$ action. The discrepancy does not affect the resulting bisimilarity. One advantage of the
weaker condition is a weaker divergence-preserving obligation in checking
divergence-preserving branching bisimulation.

It is also worth noting that the divergence preserving condition for $\approx^\vartriangle_b$
is more strict than that required in the divergence preserving relations introduced earlier in \cite{walker90}
and \cite{lohrey05}, in that the earlier works does not concern about the different equivalence classes
that passed through by a divergent run.

\begin{lemma}\label{progressingtod.p.} A strong bisimulation is a progressing branching bisimulation, which in turn is
a divergence-preserving branching bisimulation, which in turn is a branching bisimulation.
\end{lemma}
{\bf Proof.} Immediately follows from Lemma \ref{distinguishingpower}.
\hfill\qed

The idea of progressing branching bisimulation
comes from the notion of progressing weak bisimulation in \cite{MS91}, where by applying the standard
fixed-point definition as in Definition \ref{bbwed}, the notion of progressing weak bisimulation
results in an equivalence relation which
is a congruence. While this is not the case for progressing branching bisimulation, i.e.
progressing branching bisimulation equivalence defined in similar fashion is not a congruence. Here
we refrain from introducing a new equivalence,
but instead, with Lemma \ref{progressingtod.p.}, we can use progressing branching bisimulation as a tool to establish
divergence-preserving branching bisimulation,
and it turned out to suit this role very well.

\begin{proposition}\label{distinguishingpower2} $\sim\ \subseteq\ \approx^\vartriangle_b\ \subseteq\ \approx_b$.
\end{proposition}
{\bf Proof.} Immediately follows from Lemma \ref{progressingtod.p.}. \hfill\qed

\begin{proposition}\label{proposition2} Let $Id_\E=\setof{(E,E)}{E\in\E}$. Then
${\cal B}(Id_\E)$ is a divergence-preserving branching bisimulation.
\end{proposition}
{\bf Proof.} Easy to check. \hfill\qed

Proposition \ref{proposition2} can be strengthened to state that
${\cal B}(\approx^\vartriangle_b)$ is a divergence-preserving branching bisimulation, which
can be a very useful technique in establishing
divergence-preserving branching bisimulation, since the troublesome
condition about infinite $\tau$-run in ${\cal B}^\vartriangle$ is avoided. However
Proposition \ref{proposition2} is sufficient for the following development.

\begin{lemma}\label{branchingaxiom} Let $E,F\in\E$. Then $\tau.(E+F)+F\approx^\vartriangle_bE+F$.
\end{lemma}
{\bf Proof.} It is easy to verify that $(\tau.(E\!+\!F)\!+\!F,E\!+\!F)\in{\cal B}(Id_\E)$.\\
By Proposition \ref{proposition2} ${\cal B}(Id_\E)$ is a divergence-preserving branching bisimulation,
thus
$\tau.(E+F)+F\approx^\vartriangle_bE+F$ follows from Definition \ref{bbwed}.
\hfill\qed

\begin{lemma}\label{BranBiAndGuarded} Let $E,F\in\E, X\in{\cal V}$. If $E\approx^\vartriangle_bF$
then $E\,\tilde\triangleright X$ if and only if $F\,\tilde\triangleright X$.
\end{lemma}
{\bf Proof.} Straightforward from Definition \ref{DefofStrongBisim} and \ref{bbwed}.\hfill\qed

Divergence-preserving branching bisimilarity $\approx^\vartriangle_b$ is not a congruence
on $\E$. For a simplest counter example, note that
and $a.0\approx^\vartriangle_b\tau.a.0$  while
$a.0+b.0\not\approx^\vartriangle_b\tau.a.0+b.0$ when
$a, b$ are different non-$\tau$ actions. This is solved in \cite{GLS20}
by adding a rootedness condition.
\begin{definition}{\sl
Two expressions $E$ and $F$ are rooted divergence-preserving branching bisimilar, notation $E=^\vartriangle_bF$,
if the following hold:
\begin{enumerate}
\item whenever $E\arrow{a}E'$ then $F\arrow{a}F'$ with $E'\approx^\vartriangle_b F'$;
\item whenever $F\arrow{a}F'$ then $E\arrow{a}E'$ with $E'\approx^\vartriangle_b F'$;
\item $E\triangleright X$ if and only if $F\triangleright X$.\hfill
\end{enumerate}}
\end{definition}

\begin{proposition}\label{strongcongandbrancong} $\sim\ \subseteq\ =^\vartriangle_b$.
\end{proposition}
{\bf Proof.} Immediately follows from the definitions. \hfill\qed


%

%

The following theorem shows that  $=^\vartriangle_b$ is a congruence relation,
thus from now on we call it  divergence-preserving branching
congruence.

\begin{theorem}\label{theorem2.8} $=^\vartriangle_b$ is a congruence on $\E$, i.e. if $E\!=^\vartriangle_b\!F$ then
$a.E=^\vartriangle_ba.F, E+D=^\vartriangle_bF+D,D+E=^\vartriangle_bD+F,$
and $\mu X.E=^\vartriangle_b\mu X.F$
for arbitrary $a\in {\cal A}_\tau, D\in\E, X\in{\cal V}$.
\end{theorem}
{\bf Proof.} Only $\mu X.E=^\vartriangle_b\mu X.F$ needs a proof
($E\!=^\vartriangle_b\!F$ assumed), all the rest
are easy. The divergence-preserving nature of the relation
made it much harder to prove than expected. A detailed proof is presented in \cite{GLS20}.
\hfill\qed

Rob van Glabbeek et al. also proved in \cite{GLS20}
that $=^\vartriangle_b$ is the weakest congruence that implies divergence-preserving
branching bisimilarity.

We close this section by introducing two versions of the very useful up-to technique. The notion of
strong bisimulation up to $\sim$ is well known (\cite{milner}), while that of
strong bisimulation up
to $\approx^\vartriangle_b$
is new.

\begin{definition}\label{defofupto}{\sl A binary relation $R\subseteq\E\times\E$ is a {\rm strong bisimulation
up to} $\sim$ if it is symmetric and $R\subseteq{\cal S}(\sim\!R\!\sim).$


A binary relation $R\subseteq\E\times\E$ is a {\rm strong bisimulation
up-to} $\approx^\vartriangle_b$ if it is symmetric and
moreover the following hold for
all $(E,F)\in R$:
\begin{enumerate}
\item whenever $E\!\arrow{\tau}\!E'$ then there exists $F'$ such that \\
$F\!\arrow{\tau}\!F'$ and $(E',F')\in R$;
\item whenever $E\arrow{a}E'$ for $a\not=\tau$ then there exists $F'$ such that $F\arrow{a}F'$
and $(E',F')\in\,\approx^\vartriangle_b\!R\approx^\vartriangle_b$;
\item whenever $E\vartriangleright\!X$ then $F\vartriangleright\!X$.
\end{enumerate}
}\end{definition}
\begin{lemma}\label{upto} Let $R\subseteq\E\times\E$.
\begin{enumerate}
\item If $R$ is a strong bisimulation up to $\sim$, then $R\subseteq\,\sim$.
\item
If $R$ is a strong bisimulation up to $\approx^\vartriangle_b$, then $R\subseteq\,=^\vartriangle_b$.
\end{enumerate}
\end{lemma}
{\bf Proof.} 1) is proved in \cite{milner}. For 2), we first show that
$\approx^\vartriangle_bR\approx^\vartriangle_b$
is a divergence-preserving branching bisimulation. Once this is done, then
$R$ is a
strong bisimulation up to $\approx^\vartriangle_b$ implies that
$\approx^\vartriangle_b\!R\approx^\vartriangle_b \ \subseteq\ \approx^\vartriangle_b$.
Then $R\subseteq =^\vartriangle_b$ immediately follows from the definition of
strong bisimulation up to $\approx^\vartriangle_b$ and the definition of $=^\vartriangle_b$. In
the rest of the proof we show that
$\approx^\vartriangle_b\!R\approx^\vartriangle_b$ is a divergence-preserving branching bisimulation.

First it is easy to see that because $R$ is symmetric then so is $\approx^\vartriangle_bR\approx^\vartriangle_b$.
We need the following simple property of $\approx^\vartriangle_b$ which is easy to establish:
whenever $F\approx^\vartriangle_bG$ and $F\dobarrow{}F'$ then there exists $G'$ such that
$G\dobarrow{}G'$ and $F'\approx^\vartriangle_bG'$.

Suppose $E\approx^\vartriangle_bR\approx^\vartriangle_bF$, and $E\arrow{a}E'$, we shall show that
\begin{enumerate}
\item[(A)] either $a=\tau$ and there exists $F'$ such that $F\dobarrow{}F'$, $E\approx^\vartriangle_bR\approx^\vartriangle_bF'$
and $E'\approx^\vartriangle_bR\approx^\vartriangle_bF'$;
\item[(B)] or there exist $F',F''$ such that
$F\dobarrow{}F'$, $F'\arrow{a}F''$, such that $E\approx^\vartriangle_bR\approx^\vartriangle_bF'$,
$E'\approx^\vartriangle_bR\approx^\vartriangle_bF''$.
\end{enumerate}

According to the meaning of relation composition, there exist $G,H\in\E$ such that
$E\approx^\vartriangle_bG$, $(G,H)\in R$, and $H\approx^\vartriangle_bF$.
By the branching bisimulation property of $\approx^\vartriangle_b$,
for the transition $E\arrow{a}E'$, either of the following must hold:
\begin{enumerate}
\item[(a)] $a=\tau$ and there exists $G'$ such that $G\dobarrow{}G'$, $E\approx^\vartriangle_bG'$
and $E'\approx^\vartriangle_bG'$,
\item[(b)] there exist $G',G''$ such that
$G\dobarrow{}G'$, $G'\arrow{a}G''$, such that $E\approx^\vartriangle_bG'$,
$E'\approx^\vartriangle_bG''$.
\end{enumerate}
We will show that (a) implies (A), and (b) implies either (A) or (B) to fulfill the above proof obligation.

If (a) is the case, from $G\dobarrow{}G'$, according to case 1) in the definition of strong bisimulation
up to $\approx^\vartriangle_b$, there exists $H'$ such that $H\dobarrow{}H'$ and $(G',H')\in R$,
and then because $H\approx^\vartriangle_bF$, there exists $F'$ such that $F\dobarrow{}F'$ and
$H'\approx^\vartriangle_bF'$. To summarize, in this case we find $F'$, such that $F\dobarrow{}F'$, and
$E\approx^\vartriangle_bR\approx^\vartriangle_bF'$, and $E'\approx^\vartriangle_bR\approx^\vartriangle_bF'$,
that is to say (A) holds.

If (b) is the case, from $G\dobarrow{}G'$ there exists $H'$ such that $H\dobarrow{}H'$
and $(G',H')\in R$, and since $R$ is a strong bisimulation up to $\approx^\vartriangle_b$,
the transition $G'\arrow{a}G''$ implies that there exists $H''$ such that $H'\arrow{a}H''$ and
$G''\approx^\vartriangle_bR\approx^\vartriangle_bH''$ (of cause it holds when $a=\tau$).
Now because $H\approx^\vartriangle_bF$, from the move $H\dobarrow{}H'$ there
must exist $F_0$ such that $F\dobarrow{}F_0$ and $H'\approx^\vartriangle_bF_0$. Now from
$H'\arrow{a}H''$, then since $H'\approx^\vartriangle_bF_0$,
either $a=\tau$ and there exists $F'$ such that $F_0\dobarrow{}F'$
and $H'\approx^\vartriangle_bF'$ and $H''\approx^\vartriangle_bF'$, in this case
we obtain $F\dobarrow{}F'$ such that $E\approx^\vartriangle_bR\approx^\vartriangle_bF'$
and $E'\approx^\vartriangle_bR\approx^\vartriangle_bF'$, that is to say (A) holds,
or there exist $F',F''$ such that $F_0\dobarrow{}F'$, $F'\arrow{a}F''$ and
$H'\approx^\vartriangle_bF'$, $H''\approx^\vartriangle_bF''$, in this case
we obtain $F\dobarrow{}F'$ and $F'\arrow{a}F''$ such that
$E\approx^\vartriangle_bR\approx^\vartriangle_bF'$ and $E'\approx^\vartriangle_bR\approx^\vartriangle_bF''$,
that is to say (B) holds. To summarize, (b) implies either (A) or (B).

If $E\vartriangleright X$, in the same way we can show that there exists $F'$ such that
$F\dobarrow{}F'$, $E\approx^\vartriangle_bR\approx^\vartriangle_bF'$, and $F'\vartriangleright X$.

Suppose $EE_1\ldots E_i\ldots$ is an infinite $\tau$-run, we have to show that
there exists some $E_k$ on the infinite $\tau$-run and also exists $F'$ such
that $F\arrow{\tau}\dobarrow{}F'$ and $E_k\approx^\vartriangle_bR\approx^\vartriangle_bF'$.

In this case, again we can assume that there exist $G,H\in\E$ such that
$E\approx^\vartriangle_bG$, $(G,H)\in R$, and $H\approx^\vartriangle_bF$. Then
it is not difficult (could be lengthy though) to argue that there exists an
infinite $\tau$-run $GG_1\ldots G_j\ldots$, such that for each $G_j$ on the run there
exists $E_i$ on the $\tau$-run from $E$ with $E_i\approx^\vartriangle_bG_j$. Then by the
definition of strong bisimulation up to $\approx^\vartriangle_b$ there exists
an infinite $\tau$-run $HH_1\ldots H_j\ldots$ such that for each $H_j$ on the run,
$G_j$ on the corresponding position of the $\tau$-run from $G$
satisfies $(G_j,H_j)\in R$. Then because $H\approx^\vartriangle_bF$,
for this infinite $\tau$-run
from $H$ there exists a $H_k$ on the $\tau$-run from $H$ and also exists $F'$
such that $F\arrow{\tau}\dobarrow{}F'$ and $H_k\approx^\vartriangle_bF'$, and with
$H_k$ we can find $G_k$ on the $\tau$-run from $G$ such that $(G_k,H_k)\in R$,
and with $G_k$ we can find some $E_i$ on the $\tau$-run from $E$ such that
$E_i\approx^\vartriangle_bG_k$. To summarize, we find $E_i$ on
the $\tau$-run from $E$ and $F'$ such that $F\arrow{\tau}\dobarrow{}F'$ and
$E_i\approx^\vartriangle_bR\approx^\vartriangle_bF'$.
\hfill\qed

\section{The Inference System and Its Soundness}
In this section we present our inference system for $=^\vartriangle_b$ and prove its soundness.
The following is the set of axioms and rules of the
inference system, besides the rules for equational reasoning (reflexivity, symmetry, transitivity, and
substituting equal for equal):
\begin{itemize}
\item[${\rm S1}$] $E+F=F+E$
\item[${\rm S2}$] $E+(F+G)=(E+F)+G$
\item[${\rm S3}$] $E+E=E$
\item[${\rm S4}$] $E+{\bf 0}=E$
\item[${\rm B}$]  $a.(\tau.(E+F)+F)=a.(E+F)$
\item[${\rm R0}$] $\mu X.E=\mu Y.(E\{Y/X\})\mbox{}\hspace{0.5cm}\hfill(Y\!\notin\!FV(\mu X.E))$
\item[${\rm R1}$] $\mu X.E=E\{\mu X.E/X\}$
\item[${\rm R2}$] If $F\!=\!E\{F/X\}$ then $F\!=\!\mu X.E$, \\
provided $X$ is guarded in $E$
\item[${\rm R3}$] $\mu X.(X+E)=\mu X.E$
\item[${\rm R4}$] $\mu X.(\tau.(\tau.E\!+\!F)\!+\!G)\!=\!\mu X.(\tau.(E\!+\!F)\!+\!G)\hfill(E\,\tilde\triangleright X)$
\item[${\rm R5}$] $\mu X.(\tau.\mu Y.(\tau.Y\!\!+\!E)\!+\!F)\!=\!\mu X.(\tau.\mu Y.E\!+\!F)\hfill(E\,\tilde
\triangleright X)$
\item[${\rm R6}$] $\mu X.\tau.E=\tau.\mu X.E\{\tau.X/X\}$
\item[${\rm R7}$] $\mu X.(\tau.X\!\!+\!\mu Y.(\tau.Y\!\!+\!E))\!=\!\mu X.\mu Y(\tau.Y\!\!+\!E)$
\item[${\rm R8}$] $\mu X.\mu Y.(\tau.(X\!\!+\!E)\!+\!F)\!=\!\mu X.\mu Y.(\tau.(Y\!\!+\!E)\!+\!F)$
\end{itemize}

We write $\ent E=F$ if $E=F$ can be inferred from the above axioms and rules
through equational reasoning.
The aim of this section is to establish the soundness of
the inference system with respect to 
$=^\vartriangle_b$, i.e.
Theorem \ref{theorem3.1}.

${\rm S1}$-${\rm S4}$ are familiar axioms which appear in axiomatisations for all bisimulation based congruences.
It is easy to show that ${\rm S1}$-${\rm S4}$ are sound with respect to $\sim$ (\cite{milner}),
hence are also sound with respect to $=^\vartriangle_b$ here by proposition \ref{strongcongandbrancong}.

${\rm B}$ is the branching axiom which was first introduced in \cite{vonglaweij96} and proved
sound with respect to (divergence blind) branching congruence.
The following lemma proves that ${\rm B}$ is also sound with respect to $=^\vartriangle_b$.
\begin{lemma} Let $E,F\in\E, a\in Act_{\tau}$. Then $$a.(\tau.(E+F)+F)=^\vartriangle_ba.(E+F).$$
\end{lemma}
{\bf Proof.} Directly follows from Lemma \ref{branchingaxiom}.
\hfill\qed

${\rm R0}$ is the axiom of $\alpha$-conversion, which is known to be sound with respect to
$\sim$, hence also sound with respect to $=^\vartriangle_b$ because of proposition \ref{strongcongandbrancong}.

${\rm R1}$-${\rm R3}$ are three rules for recursion
introduced in \cite{milner84}, and can be proved sound with respect to $\sim$, hence
${\rm R1}$ and ${\rm R3}$ are also sound with respect to $=^\vartriangle_b$. As an equational rule with
an equality as premise, the soundness of ${\rm R2}$ with respect to $=^\vartriangle_b$ does not
immediately  follow from its soundness with respect to $\sim$, hence we need the
following lemma for the soundness of
${\rm R2}$ here.

\begin{lemma}\label{soundnessofR2for=1} If $X$ is guarded in $E$, $F=^\vartriangle_bE\{F/X\}$, then $F=^\vartriangle_b\mu X.E$.
\end{lemma}
{\bf Proof.} Construct the following relation:
$$S=\setof{(H\{F/X\},H\{\mu X.E/X\})}{H\!\in\!\E, X\mbox{\,is guarded in }H}.$$
We show that $S\cup S^{-1}$ is a strong bisimulation up to $\approx^\vartriangle_b$.
Once this is done, then since $(E\{F/X\},E\{\mu X.E/X\})\in S\cup S^{-1}$,
it follows from Lemma \ref{upto} that $E\{F/X\}=^\vartriangle_bE\{\mu X.E/X\}$,
and then $F=^\vartriangle_bE\{F/X\}=^\vartriangle_bE\{\mu X.E/X\}=^\vartriangle_b\mu X.E$.

To show that $S\cup S^{-1}$ is a strong bisimulation up to $\approx^\vartriangle_b$,
suppose $(H\{F/X\},H\{\mu X.E/X\})\!\in\!S$ and $H\{F/X\}\!\arrow{a}\!L$.\\
Then since $X$ is guarded in $H$, according to 1) of Lemma \ref{glabblemma1}
it must be that $H\arrow{a}H'$ and $L\equiv H'\{F/X\}$, thus \\$H\{\mu X.E/X\}\arrow{a}H'\{\mu X.E/X\}$.
If $a=\tau$, then $X$ must still be
guarded in $H'$, then
$(H'\{F/X\},H'\{\mu X.E/X\})\in S$.\\ If
$a\not=\tau$, then $X$ could be unguarded in $H'$. However, since $X$ is guarded in $E$,
in this case
%
$X$ is sill guarded in $H'\{E/X\}$, and
$H'\{F/X\}=^\vartriangle_bH'\{E\{F/X\}/X\}$,\\
$(H'\{E\{F/X\}/X\},H'\{E\{\mu X.E/X\}/X\})\in S\cup S^{-1}$, \\
$H'\{E\{\mu X.E/X\}/X\}=^\vartriangle_bH'\{\mu X.E/X\}$, thus\\
$(H'\{F/X\},H'\{\mu X.E/X\})\in\ \approx^\vartriangle_b(S\cup S^{-1})\approx^\vartriangle_b$.
Also since $X$ is guarded in $H$, it easily follows from 2) of Lemma \ref{glabblemma1} that
if $H\{F/X\}\vartriangleright Y$ then $H\{\mu X.E/X\}\vartriangleright Y$.

If $(H\{F/X\},H\{\mu X.E/X\})\!\in\!S^{-1}$, in the same way we can show that the conditions
1),2),3)
in Definition \ref{defofupto} are satisfied. So $S\cup S^{-1}$ is a strong bisimulation up to
$\approx^\vartriangle_b$.
\hfill\qed


${\rm R4}$ as an axiom was first introduced in \cite{glabb96} for eliminating $\tau$'s in front of
unguarded occurrences of bound variables.
For $=^\vartriangle_b$ however,
${\rm R4}$ alone is not enough to
eliminate all such $\tau$'s. So here we introduce ${\rm R5}$ to work together with ${\rm R4}$.
Intuitively, ${\rm R5}$ means that, the presence of unguarded occurrences
of $X$ in $E$ implies a $\tau$-circle going back to the recursion
represented by $X$, thus the inner $\tau$-loop for the recursion represented by
$Y$ can be eliminated without changing the divergent behaviour of the expression.
The following lemma states the soundness of ${\rm R4}$ and ${\rm R5}$.
The proof uses the technique of progressing branching bisimulation.
\begin{lemma}\label{progtechnique} Let $E,F,G\in\E, X,Y\in\V$, $E\,\tilde\triangleright X$, then
\begin{enumerate}
\item $\mu X.(\tau.(\tau.E+F)+G)=^\vartriangle_b\mu X.(\tau.(E+F)+G)$;
\item
$\mu X.(\tau.\mu Y.(\tau.Y+E)+F)=^\vartriangle_b\mu X.(\tau.\mu Y.E+F).$
\end{enumerate}
\end{lemma}
{\bf Proof}. To prove 1), let $L=\mu X.(\tau.(\tau.E+F)+G)$ and \\
$R=\mu X.(\tau.(E+F)+G)$, and let $S$ be the following relation:
$$\begin{array}{l}
\setof{(H\{L/X\},H\{R/X\})}{H\in\E}\cup\\
\{(E\{L/X\},E\{R/X\}+F\{R/X\}),\\
(\tau.E\{L/X\}+F\{L/X\},E\{R/X\}+F\{R/X\})\}.
\end{array}
$$
We show that $S\cup S^{-1}$ is a progressing branching bisimulation.

By Lemma \ref{glabblemma3}, $L\arrow{a}L'$ if and only if
\begin{enumerate}
\item either $G\arrow{a}G'$ and $L'\equiv G'\{L/X\}$,
\item or $a=\tau$ and $L'\equiv\tau.E\{L/X\}+F\{L/X\}$.
\end{enumerate}
For the same reason $R\arrow{a}R'$ if and only if
\begin{enumerate}
\item either $G\arrow{a}G'$ and $R'\equiv G'\{R/X\}$,
\item or $a=\tau$ and $R'\equiv E\{R/X\}+F\{R/X\}$.
\end{enumerate}
Because $E\,\tilde\triangleright X$, together with $L\arrow{\tau}\tau.E\{L/X\}+F\{L/X\}$
and $R\arrow{\tau}E\{R/X\}+F\{R/X\}$, the following follows from Lemma \ref{glabblemma2}:
\begin{enumerate}
\item $E\{L/X\}\arrow{\tau}\dobarrow{}\tau.E\{L/X\}+F\{L/X\}$, and
\item $E\{R/X\}\arrow{\tau}\dobarrow{}E\{R/X\}+F\{R/X\}$.
\end{enumerate}
From these observation, it is easy to check that $S\cup S^{-1}$ is a progressing branching bisimulation, hence
a div.-pres. branching bisimulation by Lemma \ref{progressingtod.p.}, and
$S\cup S^{-1}\subseteq\,\approx^\vartriangle_b$.
Moreover, whenever $L\arrow{a}L'$ then there exists $R'$ such that $R\arrow{a}R'$ with $(L',R')\in S$,
and whenever $R\arrow{a}R'$ then there exists $L'$ such that $L\arrow{a}L'$ with $(R',L')\in S^{-1}$,
and $L\vartriangleright Y$ if and only if $R\vartriangleright Y$ for any $Y\in{\cal V}$,
thus $L=^\vartriangle_b R$.

To prove 2), let
$L_X\!=\!\mu X.(\tau\!.\mu Y.(\tau.Y\!+\!E)\!+\!F), \\
L_Y\!=\!\mu Y.(\tau.Y\!+\!E\{L_X/X\}),
R_X\!=\!\mu X.(\tau\!.\mu Y.E\!+\!F),$ and \\
$R_Y=\mu Y.E\{R_X/X\}$, and let $S$ be the following relation:
$$\begin{array}{l}
\setof{(H\{L_X/X\},H\{R_X/X\})}{H\in\E}\cup\\
\setof{(H\{L_X/X,L_Y/Y\},H\{R_X/X,R_Y/Y\})}{H\in\E}.
\end{array}
$$
We show that $S\cup S^{-1}$ is a progressing branching bisimulation.

By Lemma \ref{glabblemma3}, $L_X\arrow{a}L'$ if and only if
\begin{enumerate}
\item either $F\arrow{a}F'$ and $L'\equiv F'\{L_X/X\}$,
\item or $a=\tau$ and $L'$ is just $L_Y$.
\end{enumerate}
And $L_Y\arrow{a}L'$ if and only if
\begin{enumerate}
\item either $a=\tau$ and $L'$ is just $L_Y$,
\item or $E\arrow{a}E'$ and $L'\equiv E'\{L_X/X,L_Y/Y\}$,
\item or $L_X\arrow{a}L'$, in which case either $F\arrow{a}F'$ and $L'\equiv F'\{L_X/X\}$
or $a=\tau$ and $L'$ is $L_Y$.
\end{enumerate}
For the same reason $R\arrow{a}R'$ if and only if
\begin{enumerate}
\item either $F\arrow{a}F'$ and $R'\equiv F'\{R_X/X\}$,
\item or $a=\tau$ and $R'$ is just $R_Y$.
\end{enumerate}
And $R_Y\arrow{a}R'$ if and only if
\begin{enumerate}
\item either  $E\arrow{a}E'$ and $R'\equiv E'\{L_X/X,L_Y/Y\}$,
\item or $R_X\arrow{a}R'$, in which case either $F\arrow{a}F'$ and $R'\equiv F'\{R_X/X\}$
or $a=\tau$ and $R'$ is $R_Y$.
\end{enumerate}
Because $E\,\tilde\triangleright X$, together with
$R_X\arrow{\tau}R_Y$, it follows from Lemma \ref{glabblemma2} that
$E\{R_X/X,R_Y/Y\}\arrow{\tau}\dobarrow{}R_Y$, thus $R_Y\arrow{\tau}\dobarrow{}R_Y$.
From these observation, it is easy to check that
$S\cup S^{-1}$ is a progressing branching bisimulation, hence
a div.-pres. branching bisimulation, and
$S\cup S^{-1}\subseteq\,\approx^\vartriangle_b$.
Moreover, whenever $L_X\arrow{a}L'$ then there exists $R'$ such that $R_X\arrow{a}R'$ with $(L',R')\in S$,
and whenever $R_X\arrow{a}R'$ then there exists $L'$ such that $L_X\arrow{a}L'$ with $(R',L')\in S^{-1}$,
and $L_X\vartriangleright Z$ if and only if $R_X\vartriangleright Z$ for any $Z\in{\cal V}$,
thus $L_X=^\vartriangle_b R_X$.
\hfill\qed

A key idea proposed by Milner in \cite{milner89} for equational
axiomatisation of this kind is to transform arbitrary expressions
into guarded ones, i.e. expressions in which every recursive subexpression
is a guarded recursion, so that the fixed point induction rule ${\rm R2}$
can be applied to derive equality of semantically equivalent expressions.
Determined by its divergence preserving nature, a
major difficulty in an axiomatisation for $=^\vartriangle_b$ is that
unguarded recursions cannot be eliminated completely
like in the axiomatisation for $=_b$.
Thus, in order to use the full power of the fixed point induction rule ${\rm R2}$,
careful manipulation of unguarded recursions is called for,
and axioms ${\rm R6}$, ${\rm R7}$, and ${\rm R8}$ are exactly for that
purpose. The intuition for ${\rm R6}$ is that, the left hand side expression
is a recursion starting with a $\tau$, while the right hand side expression
always perform a $\tau$ before recursion, so they should be doing the same thing.
${\rm R7}$ roughly says that double loop is the same as a single loop, which
intuitively makes sense.
The intuition for ${\rm R8}$ is that here the successive recursion for
$X$ and $Y$ effectively defines $X$ and $Y$ as the same behaviour,
thus interchanging the two variables should not affect the overall behaviour.
The following lemma states the soundness of ${\rm B6}$, ${\rm R7}$,
and ${\rm R8}$ with respect to $=^\vartriangle_b$. Surprisingly it turned out that
these axioms are even sound with respect to the stronger congruence $\sim$.

\begin{lemma}\label{muT} Let $E,F\in\E, X,Y\in {\cal V}$. Then
\begin{enumerate}
\item $\mu X.\tau.E\sim\tau.\mu X.E\{\tau.X/X\};$
\item $\mu X.(\tau.X\!+\mu Y.(\tau.Y\!+E))\sim\mu X.\mu Y(\tau.Y\!+E);$
\item $\mu X.\mu Y.(\tau.(X\!+\!E)\!+\!F)\sim\mu X\!.\mu Y\!.(\tau\!.(Y\!\!+\!E)\!+\!F).$
\end{enumerate}
\end{lemma}
{\bf Proof.} To prove 1), let 
$$S=\setof{(H\{\mu X.\tau.E/X\},H\{\tau.\mu X.E\{\tau.X/X\}/X\}}{H\in\E}.$$
Then $S\cup S^{-1}$ is a strong bisimulation
up to $\sim$, thus $S\subseteq\,\sim$. \\
Take $X$ as $H$, then
$(\mu X.\tau.E,\tau.\mu X.E\{\tau.X/X\})\in S$,
thus\\
$\mu X.\tau.E\sim\tau.\mu X.E\{\tau.X/X\}$.

To prove 2), let
$$S=\setof{(H\{L_X/X,L_Y/Y\},H\{R_X/X,R_Y/Y\})}{H\in\E}
$$
where\\
$\begin{array}{ll}
\!\!\!L_X\!=\!\mu X.(\tau.X\!\!+\!\mu Y.(\tau.Y\!\!+\!E)),& \!\!\!\!L_Y\!=\!\mu Y.(\tau.Y\!\!+\!E\{L_X/X\}),\\
\!\!\!R_X\!=\!\mu X.\mu Y.(\tau.Y\!+E),& \!\!\!\!R_Y\!=\!\mu Y.(\tau.Y\!\!+\!E\{R_X/X\}).
\end{array}
$
Note that $R_X\sim R_Y$, then
$S\cup S^{-1}$ is a strong bisimulation up-to $\sim$, thus
$S\subseteq\ \sim$ follows from Lemma \ref{upto}.
Take $X$ as $H$, we have
$(L_X,R_X)\in S$, hence \\
$\mu X.(\tau.X\!+\mu Y.(\tau.Y\!+E))\sim\mu X.\mu Y(\tau.Y\!+E)$.

To prove 3), let
$$S=\setof{(H\{L_X/X,L_Y/Y\},H\{R_X/X,R_Y/Y\})}{H\in\E}
$$
where\\
$\begin{array}{ll}
\!\!\!L_X\!=\!\mu X.\mu Y.(\tau.(X\!\!+\!E)+F),& \\
\!\!\!L_Y\!=\!\mu Y.(\tau.(L_X\!\!+\!E\{L_X/X\})+F\{L_X/X\}),\\
\!\!\!R_X\!=\!\mu X.\mu Y.(\tau.(Y\!+E)+F),& \\
\!\!\!R_Y\!=\!\mu Y.(\tau.(Y\!\!+\!E\{R_X/X\})+F\{R_X/X\}).
\end{array}
$\\
Note that $R_X\sim R_Y$, then
$S\cup S^{-1}$ is a strong bisimulation up-to $\sim$, thus
$S\subseteq\ \sim$. Take $X$ as $H$ to obtain $(L_X,R_X)\in S$,
hence $\mu X.\mu Y.(\tau.(X\!+\!E)\!+\!F)\sim\mu X\!.\mu Y\!.(\tau\!.(Y\!\!+\!E)\!+\!F).$
\hfill\qed

With these lemmas, finally
we have the following soundness theorem for
the inference system with respect to $=^\vartriangle_b$.
\begin{theorem}\label{theorem3.1} For $E,F\in\E$, if $\ent E=F$ then $E=^\vartriangle_b F$.
\end{theorem}
{\bf Proof}.
Since $=^\vartriangle_b$ is an equivalence relation, equational reasoning preserves soundness.
Also by Theorem \ref{theorem2.8} $=^\vartriangle_b$ is a congruence, thus the inference rule of
substituting equal for equal preserves soundness.
We also know the soundness of ${\rm S1}$-${\rm S4}$, ${\rm B}$, ${\rm R0}$-${\rm R7}$.
Thus if $\ent E=F$ then $E=^\vartriangle_bF$.
\hfill\qed

Before closing this section we prove two useful derived rules.
Milner's inference system in \cite{milner89} included the following axiom ${\rm T1}$:
$a.\tau.E=a.E.$
The following
theorem shows that with ${\rm B}$, ${\rm T1}$ can be derived from the present axiomatisation.
\begin{theorem}\label{T1asDerivedRule} Let $E\in\E$. Then: $${\rm T1}\ \ \ent a.\tau.E=a.E.$$
\end{theorem}
{\bf Proof}. $\ent a.\tau.E=a.(\tau.(E+\nil)+\nil)=a.(E+\nil)=a.E$.
\hfill\qed
Thus ${\rm T1}$ can be used as a derived rule in the inference system.

The next derived rule is ${\rm D0}$ as
stated in Theorem \ref{derivedruleD0}. To prove it
we need the following lemma.
\begin{lemma}\label{Summand} Let $E,E'\in\E, X\in\V$.
\begin{enumerate}
\item If $E\arrow{a}E'$ then $\ent E=E+a.E'$;
\item If $E\vartriangleright X$ then $\ent E=E+X$.
\end{enumerate}
\end{lemma}
{\bf Proof.} By straightforward induction
on the set of
rules defining the relations $\arrow{}$ and $\vartriangleright$ (Definition \ref{defoftrans}).
\hfill\qed

\begin{theorem}\label{derivedruleD0} For $E,F\in\E, X\in\V$, the following holds:\\
${\rm D0}$ If $E\,\tilde\triangleright X$ then
$\ent \mu X.(\tau.E+F)=\mu X.(\tau.(X+E)+F)$.
\end{theorem}
{\bf Proof.}
Since $E\,\tilde\triangleright X$ is $E\dobarrow{}\vartriangleright\!X$, i.e.
there exist $E_1,\ldots,E_n$ such that
$E_{i-1}\!\arrow{\tau}\!E_{i}$ 
for $i=1,\ldots,n$ ($E_0\equiv E$) and
$E_n\!\vartriangleright\!X$.
By Lemma \ref{Summand}  $\ent E_{i-1}=E_{i-1}+\tau.E_{i}$
for $i=1,\ldots,n$ and $\ent E_n=E_n+X$.
Then we have
the following proof:
\begin{enumerate}
\item[$\ent$] $\mu X.(\tau.E+F)=\mu X.(\tau.E_0+F)$
\item[$=$] $\mu X.(\tau.(E_0+\tau.E_{1})+F)$\hfill Lemma \ref{Summand}
\item[$=$] $\mu X.(\tau.(E_0+E_{1})+F)\hfill E_{1}\,\tilde\triangleright X,{\rm R4}$
\item[$=$] $\mu X.(\tau.(E_0+\ldots+E_n)+F)$\hfill repeating above step
\item[$=$] $\mu X.(\tau.(E_0+\ldots+E_n+X)+F)$\hfill Lemma \ref{Summand}
\item[$=$] $\mu X.(\tau.(X\!+\!E_0\!+\!\ldots\!+\!E_{n-1}\!+\!E_n)\!+\!F)\hfill {\rm S1,S2}$
\item[$=$] $\mu X.(\tau.(X\!+\!E_0\!+\!\ldots\!+\!E_{n-1}\!+\!\tau.E_n)\!+\!F)\hfill {\rm R4}$
\item[$=$] $\mu X.(\tau.(X\!+E_0+\ldots+E_{n-1})+F)$\hfill Lemma \ref{Summand}
\item[$=$] $\mu X.(\tau.(X+E_0)+F)$\hfill repeating above step
\item[$=$] $\mu X.(\tau.(X+E)+F)$\hfill\qed
\end{enumerate}


\section{Loop Operator and Standard Sum}

The {\em loop operator} $\tau^*\underline{\ }$ applied on an expression $E$
obtains the loop expression $\tau^*E$ which can choose to perform
$\tau$ without changing its state or choose to perform actions of $E$.
In \cite{lohrey05},
the loop expression $\tau^*E$ was included
into the basic syntax of expressions (where the notation used is $\Delta(E)$), and played
a key role in defining (guarded) standard forms for all expressions.
The syntax of basic expressions of this paper does not include the loop operator $\tau^*\underline{\ }$,
we introduce the following definition instead.
\begin{definition}\label{defofloop}{\sl
Let $E$ be an expression,
define the loop expression $\tau^*E$ as $\mu X.(\tau X+E)$ where $X\notin FV(E)$.}
\end{definition}

That is we use $\tau^*E$ as an abbreviation,
then we can reuse the generalized notion of guarded expression
introduced in \cite{lohrey05}. Excluding the loop operator
from the basic syntax helps to simplify the
semantic theory (so there is no  need to define particularly
the operational semantics of $\tau^*E$, and to study
the congruence property concerning $\tau^*E$ etc.) and to keep a small set of core axioms of the inference system.

In this section we first prove some derived rules relating to
the loop operator. These derived rules, together with
${\rm T1}$ (Theorem \ref{T1asDerivedRule}) and
${\rm D0}$ (Theorem \ref{derivedruleD0}),
will be used in the standardization process.

\begin{theorem} The following equalities can be derived from the inference system.
\begin{itemize}
\item[${\rm D1}$] $\ent \tau^*E=\tau.(\tau^*E)+E$;
\item[${\rm D2}$] $\ent \tau^*E=\tau^*E+E$;
\item[${\rm D3}$] $\ent \mu X.(\tau.(X+E)+F)=\mu X.(\tau.\tau^*(E+F)+F)$;
\item[${\rm D4}$] $\ent \mu X\!.(\tau\!.(\!X\!+\!E)\!+\!\tau\!.(\!X\!+\!F)\!+\!G)\!=\!
\mu X\!.(\tau\!.(\!X\!+\!E\!+\!F)\!+\!G)$;
\item[${\rm D5}$] $\ent \tau^*(\tau.\tau^*(E+F)+F)=\tau.\tau^*(E+F)+F$;
\item[${\rm D6}$] $\ent \tau^*(\tau^*E)=\tau^*E$.
\end{itemize}
\end{theorem}
{\bf Proof}.
${\rm D1}$ is a special instance of  ${\rm R1}$.

The following is a proof of ${\rm D2}$:
\begin{itemize}
\item[$\ent$] $\tau^*E=\tau.\tau^*E+E\hfill{\rm D1}$
\item[$=$] $\tau.\tau^*E+E+E\hfill{\rm S3}$
\item[$=$] $\tau^*E+E\hfill{\rm D1}$
\end{itemize}

The following is a proof of ${\rm D3}$, where $Y$ is a variable which
occurs neither free in $E$ nor in $F$:
\begin{itemize}
\item[$\ent$] $\mu X.(\tau.(X+E)+F)$
\item[$=$] $\mu X.\mu Y.(\tau.(X+E)+F)\hfill{\rm R1}$
\item[$=$] $\mu X.\mu Y.(\tau.(Y+E)+F)\hfill{\rm R8}$
\item[$=$] $\mu X.\mu Y.(\tau.(\tau.Y+E)+F)\hfill{\rm R4}$
\item[$=$] $\mu X.(\tau\!.(\tau\!.\mu Y\!.(\tau\!.(\tau\!.Y\!\!+\!E)\!+\!F)\!+\!E)\!+\!F)\hfill{\rm R1}$
\item[$=$] $\mu X.(\tau.(\mu Y.\tau.(\tau.(Y\!+\!E)\!+\!F)\!+\!E)\!+\!F)\hfill{\rm R6}$
\item[$=$] $\mu X.(\tau.(\mu Y.\tau.(Y+E+F)+E)+F)\hfill{\rm R4}$
\item[$=$] $\mu X.(\tau.(\tau.\mu Y.(\tau.Y\!+\!E\!+\!F)\!+\!E)\!+\!F)\hfill{\rm R6}$
\item[$=$] $\mu X.(\tau.(\tau.\tau^*(E\!+\!F)\!+\!E)\!+\!F)$\hfill Definition \ref{defofloop}
\item[$=$] $\mu X.(\tau\!.(\tau\!.(\tau^*(E\!+\!F)\!+\!E\!+\!F)\!+\!E)\!+\!F)\hfill{\rm D2}$
\item[$=$] $\mu X.(\tau.(\tau^*(E+F)+E+F)+F)\hfill{\rm B}$
\item[$=$] $\mu X.(\tau.\tau^*(E+F)+F)\hfill{\rm D2}$
\end{itemize}

The following is a proof of ${\rm D4}$, where $Y$ is a variable which
does not occur free in $E,F,G$:
\begin{itemize}
\item[$\ent$] $\mu X.(\tau.(X+E)+\tau.(X+F)+G)$
\item[$=$] $\mu X.(\tau.\tau^*(E\!+\!\tau.(X\!\!+\!F)\!+\!G)\!+\!\tau.(X\!\!+\!F)\!+\!G)\hfill{\rm D3}$
\item[$=$] $\mu X.(\tau.\mu Y.(\tau.Y\!\!+\!E\!+\!\tau.(X\!\!+\!F)\!+\!G)\!+\!\tau.(X\!\!+\!F)\!+\!G)$
\item[$=$] $\mu X.(\tau.\mu Y.(E\!+\!\tau.(X\!\!+\!F)\!+\!G)\!+\!\tau.(X\!\!+\!F)\!+\!G)\hfill{\rm R5}$
\item[$=$] $\mu X.(\tau.(E\!+\!\tau.(X\!\!+\!F)\!+\!G)\!+\!\tau.(X\!\!+\!F)\!+\!G)\hfill{\rm R1}$
\item[$=$] $\mu X.(\tau.(E\!+\!(X\!\!+\!F)\!+\!G)\!+\!\tau.(X\!\!+\!F)\!+\!G)\hfill{\rm R4,S1}$
\item[$=$] $\mu X.(\tau.(X\!\!+\!E\!+\!F\!+\!G)\!+\!\tau.(X\!\!+\!F)\!+\!G)\hfill{\rm S1,S2}$
\item[$=$] $\mu X.(\tau.(X\!\!+\!F)\!+\!\tau.(X\!\!+\!E\!+\!F\!+\!G)\!+\!G)\hfill{\rm S1}$
\item[$=$] $\mu X.(\tau.\tau^*(F\!+\!\tau.(X\!\!+\!E\!+\!F\!+\!G)\!+\!G)$\\
$\mbox{}\hspace{1cm} \ \ \ \ \ \ \ +\tau.(X\!\!+\!E\!+\!F\!+\!G)\!+\!G)\hfill{\rm D3}$
\item[$=$] $\mu X.(\tau.(F\!+\!\tau.(X\!\!+\!E\!+\!F\!+\!G)\!+\!G)$\\
$\mbox{}\ \ \ \ \ \ \ \ \ \ \ \ +\tau.(X\!\!+\!E\!+\!F\!+\!G)\!+\!G)\hfill{\rm R5,R1}$
\item[$=$] $\mu X.(\tau.(F\!+\!(X\!\!+\!E\!+\!F\!+\!G)\!+\!G)$\\
$\mbox{}\ \ \ \ \ \ \ \ \ \ +\tau.(X\!\!+\!E\!+\!F\!+\!G)\!+\!G)\hfill{\rm R4,S1}$
\item[$=$] $\mu X.(\tau.(X\!+\!E\!+\!F\!+\!G)\!+\!\tau.(X\!+\!E\!+\!F\!+\!G)\!+\!G)\hfill{\rm S1\mbox{-}S3}$
\item[$=$] $\mu X.(\tau.(X+E+F+G)+G)\hfill{\rm S3}$
\item[$=$] $\mu X.(\tau.\tau^*(E+F+G)+G)\hfill{\rm D3}$
\item[$=$] $\mu X.(\tau.(X+E+F)+G)\hfill{\rm D3}$
\end{itemize}

The following is a proof of ${\rm D5}$, where $X,Y$ are different variables which
occur neither free in $E$ nor in $F$:
\begin{itemize}
\item[$\ent$] $\tau^*(\tau.\tau^*(E+F)+F)$
\item[$=$] $\mu X.(\tau.X+\tau.\tau^*(E+F)+F)$\hfill Definition \ref{defofloop}
\item[$=$] $\mu X.(\tau.X+\mu Y.(\tau.\tau^*(E+F)+F))\hfill{\rm R1}$
\item[$=$] $\mu X.(\tau.X+\mu Y.(\tau.(Y+E)+F))\hfill{\rm D3}$
\item[$=$] $\mu X.(\tau.X+\mu Y.(\tau.Y+\tau.(Y+E)+F))\hfill{\rm D4,S3}$
\item[$=$] $\mu X.\mu Y.(\tau.Y+\tau.(Y+E)+F)\hfill{\rm R7}$
\item[$=$] $\mu X.\mu Y.(\tau.(Y+E)+F)\hfill{\rm D4,S3}$
\item[$=$] $\mu X.\mu Y.(\tau.\tau^*(E+F)+F)\hfill{\rm D3}$
\item[$=$] $\mu X.(\tau.\tau^*(E+F)+F)\hfill{\rm R1}$
\item[$=$] $\tau.\tau^*(E+F)+E\hfill{\rm R1}$
\end{itemize}


${\rm D6}$ can be derived from ${\rm D1}$ and ${\rm D5}$ as follows:
\begin{itemize}
\item[$\ent$] $\tau^*(\tau^*E)=\tau^*(\tau.\tau^*E+E)\hfill{\rm D1}$
\item[$=$] $\tau^*(\tau.\tau^*(\nil+E)+E)\hfill{\rm S4}$
\item[$=$] $\tau.\tau^*(\nil+E)+E)\hfill{\rm D5}$
\item[$=$] $\tau.\tau^*E+E=\tau^*E$\hfill\qed
\end{itemize}


\begin{definition}{\sl
A {\em guarded recursion} is an expression of the form $\mu X.E$ where
$X$ is guarded in  $E$. 

A {\em guarded expression} is an expression in which every subexpression of
the form $\mu X.E$
is either a loop expression or a guarded recursion.

Let $E\in\E,X\in\V$, $X$ is said {\em fully exposed in} $E$ if whenever
an unguarded occurrence of $X$ in $E$ is within a subexpression $\mu Y.F$
of $E$, then $\mu Y.F$ is a loop expression.
}
\end{definition}

The main purpose of this section is to prove the following standardization result: every expression can be proven equivalent
to a guarded expression of some
form. In the standardization process, the property that whether
a variable $X$ is
fully exposed in an expression $E$ is very important. If $X$ is not fully exposed
in a guarded expression, then there is an equivalent guarded expression in which
$X$ is fully exposed.
For example, the occurrence
$X$ in $E\equiv \tau.\mu Y.(\tau.X+a.Y)$ is unguarded, and since it is within
a recursion which is not a loop expression, $X$ is not fully exposed in $E$. Applying ${\rm R1}$ to unfold the recursion,
we obtain $\ent E=\tau.(\tau.X+a.\mu Y.(\tau.X+a.Y))$, and $X$ becomes fully exposed on the right hand side
(suppose $a\not=\tau$). The following lemma shows that this is generally true.

\begin{lemma}\label{FullyExposed} Let $E\in\E, X\in\V$. If $E$ is a guarded expression, then there exists
a guarded expression $E'$
such that $X$ is fully exposed in $E'$ and $\ent E=E'$.
\end{lemma}
{\bf Proof.}
The proof is by induction on the structure of $E$.
Here we look at the only non trivial case where
$E$ is a recursion $\mu Y.F$. Since
$\mu Y.F$ is a guarded expression, either it is a loop expression
so $F\equiv\tau.Y+F_1$ where $F_1$ is a guarded expression,
or $Y$ is guarded in $F$ which itself is a guarded expression. In the first case,
by the induction hypothesis there exists a guarded expression $F_1'$ such
that $X$ is fully exposed in $F_1'$ and
$\ent F_1=F_1'$. Then
$\mu Y.(\tau.Y+F_1')$ (which is a loop expression) is guarded and $X$ is fully exposed in
$\mu Y.(\tau.Y\!+F_1')$ and $\ent\!\mu Y.F\!=\!\mu Y.(\tau.Y\!+F_1')$.
In the second case,
by the induction hypothesis there exists a guarded expression
$F'$ such that $X$ is fully exposed in $F'$ and
$\ent F=F'$. Since $Y$ is guarded in $F$ and $F=^\vartriangle_bF'$ (follows from
$\ent F=F'$), $Y$ must be guarded also in $F'$. Then $X$ is fully exposed in $F'$
implies that $X$ is fully exposed in $F'\{\mu Y.F'/Y\}$ since those unexposed unguarded
occurrences of $X$ in $\mu Y.F'$ becomes guarded occurrences in $F'\{\mu Y.F'/Y\}$. Clearly $F'\{\mu Y.F'/Y\}$
is guarded since  $F'$ is guarded. Now $\ent \mu Y.F'=F'\{\mu Y.F'/Y\}$,
thus $\ent\mu Y.F=F'\{\mu Y.F'/Y\}$
and $F'\{\mu Y.F'/Y\}$ is the $E'$ we need in this case.
\hfill\qed



\begin{lemma}\label{Exposed} Let $E,F\in\E$, $X\in\V$.
If $E$ is guarded,
$E\,\tilde\triangleright X$, and
$X$ is fully exposed in $E$, then there exists $E_1$ such that $X$
is guarded in $E_1$, and
$\ent \mu X.(\tau.E\!+\!F)\!=\!\mu X.(\tau.(X\!+\!E_1)\!+\!F).$
\end{lemma}
{\bf Proof.} The proof is by induction on the
structure of $E$.

If $E$ is $\nil$, then $E\!\not\!\tilde\triangleright X$, and in this case the claim holds vacuously.

If $E$ is a variable, then since $E\,\tilde\triangleright X$ the variable must be $X$,
we can take $\nil$ as $E_1$, then $X$ is guarded in $E_1$ and the equality holds and $E_1$ is a guarded expression.

If $E$ is a prefix form $a.E'$, since $E\,\tilde\triangleright X$, it must be
that $a=\tau$ and $E'\,\tilde\triangleright X$. Since $X$ is fully exposed in $E$,
$X$ must be fully exposed in $E'$. Then
\begin{itemize}
\item[$\ent$] $\mu X.(\tau.E\!+\!F)=\mu X.(\tau.\tau.E'+F)$
\item[$=$] $\mu X.(\tau.E'+F)\hfill{\rm T1}$
\item[$=$] $\mu X.(\tau.(X\!+\!E_1)\!+\!F)\hfill\mbox{ind. hyp. on }E'$
\end{itemize}
where $E_1$ is obtained by applying induction hypothesis on $E'$, such that $E_1$
is a guarded expression, and
$X$ is guarded in $E_1$.

If $E$ is $E'+E''$, then $E\,\tilde\triangleright X$ and $X$ is fully exposed in $E$
implies that $X$ is fully exposed in $E'$ and $E''$, and either
$E'\,\tilde\triangleright X$ or $E''\,\tilde\triangleright X$ or both.
If $E'\,\tilde\triangleright X$ and $X$ is fully exposed in $E'$, then
\begin{itemize}
\item[$\ent$] $\mu X.(\tau.E+F)=\mu X.(\tau.(X+E)+F)\hfill{\rm D0}$
\item[$=$] $\mu X.(\tau.(X+E'+E'')+F)$
\item[$=$] $\mu X.(\tau.(X+E')+\tau(X+E'')+F)\hfill{\rm D4}$
\item[$=$] $\mu X.(\tau.E'+\tau.(X+E'')+F)\hfill{\rm D0}$
\item[$=$] $\mu X.(\tau.(X+E_1')+\tau.(X+E'')+F)\hfill\mbox{ind. hyp. on }E'$
\item[$=$] $\mu X.(\tau.(X+E'_1+E'')+F)\hfill{\rm D4}$
\end{itemize}
%
where $E_1'$ is a guarded expression obtained by applying induction hypothesis on $E'$, and
$X$ is guarded in $E_1'$. If $X$ is guarded in $E''$ then $E_1'+E''$ is the final $E_1$ which is
a guarded expression and
in which $X$ is guarded, otherwise we can do the same routine again on $E''$.

If $E$ is a recursion $\mu Y.H$, since $E\,\tilde\triangleright X$
and $X$ is fully exposed in $E$, $\mu Y.H$ must be a loop expression, that is
$\mu Y.H\equiv\tau^*E'$,
and $E'\,\tilde\triangleright X$, and $X$ is fully exposed in
$E'$.
Then:
\begin{itemize}
\item[$\ent$] $\mu X.(\tau.E+F)=\mu X.(\tau.\tau^*E'+F)$
\item[$=$] $\mu X.(\tau.E'+F)\hfill{\rm R5,R1}$
\item[$=$] $\mu X.(\tau.(X+E_1')+F)\hfill\mbox{ind. hyp. on }E'$
\end{itemize}
%
where $E_1'$ is a guarded expression obtained by applying induction hypothesis on $E'$, and
$X$ is guarded in $E_1'$.
\hfill\qed

Let $S=\{E_1,\ldots,E_n\}$ be a finite set of expressions, we write $\Sigma S$ and
$\Sigma_{i=1}^nE_i$ as abbreviations for
$E_1+\ldots+E_n$. The use of such notations is justified by the axioms ${\rm S1}$-${\rm S4}$.
\begin{definition}\label{defofsimplesum}{\sl
A {\em standard sum} is an expression
of the form $\Sigma_{i=1}^na_i.E_i+\Sigma_{j=1}^mW_j$ where $E_i$ is a guarded expression for $i=1,\ldots,n$
and $W_j\in{\cal V}$ for $j=1,\ldots,m$. 

A {\em simple standard sum} (or a simple sum) is a standard sum of the form
$\Sigma_{i=1}^na_i.X_i+\Sigma_{j=1}^mW_j$
where $X_i$'s and $W_j$'s are all variables.
}
\end{definition}

At last we can state and prove the standardization theorem.

\begin{theorem}\label{guardedness}
Every expression can be proven equivalent to a standard sum.
\end{theorem}
{\bf Proof.} 
We prove by structural induction that for every $E\in\E$ there is a standard sum
$E'$ such that $\ent E=E'$.

Here we only look at the case where
$E$ is a recursion $\mu X.F$, all other cases are simple.
By the induction hypothesis 
there exist guarded
expressions $E_1,\ldots,E_n$ and variables $W_1,\ldots,W_m$ such that
$\ent F\!=\!\Sigma_{i=1}^na_i.E_i\!+\!\Sigma_{j=1}^mW_j$.
Let 
$S$ be the set of summands in $\Sigma_{i=1}^na_i.E_i+\Sigma_{j=1}^mW_j$,
then the elements of $S$
can be divided into four groups:
\begin{enumerate}
\item those of the form $a.H$ where $X$ occurs unguarded in $a.H$;
\item those of the form $a.H$ where $X$ is guarded in $a.H$;
\item those variable $W$ which is not $X$;
\item $X$ (if it is one of the summands).
\end{enumerate}
Let $G$ be the sum of all expressions in the second and third groups above,
then $G$ is a standard guarded sum and $X$ is guarded in $G$, and
let $\tau.F_1,\ldots,\tau.F_k$
be the expressions in the first group (since $X$ occurs unguarded in these summands, the prefix must be $\tau$), then
$\ent\mu X.F=\mu X.(\Sigma_{i=1}^k\tau.F_i+G)$
(using ${\rm R3}$ to eliminate
$X$ in the fourth group if needed).
Since each $F_i$ is  a guarded expression, by Lemma \ref{FullyExposed}
there is a guarded expression $F'_i$ s. t. $X$ is fully exposed in $F'_i$ and
$\ent F_i=F'_i$, thus $\ent\mu X.F=\mu X.(\Sigma_{i=1}^k\tau.F'_i+G)$.
Also note that since the $\tau.F_i$'s are from the first group, thus $X$ occurs unguarded in $F_i$,
so $F_i\,\tilde\triangleright X$ (Lemma \ref{glabblemma2}). Then $F_i'\,\tilde\triangleright X$ follows
from Lemma \ref{BranBiAndGuarded}.
Now apply Lemma \ref{Exposed} $k$ times we find guarded expressions $F''_1,\ldots,F''_k$
such that $X$ is guarded in all $F''_i$'s and\\
$\ent\mu X.(\Sigma_{i=1}^k\tau.F'_i+G)=\mu X.(\Sigma_{i=1}^k\tau.(X+F''_i)+G)$. Then
\begin{enumerate}
\item[$\ent$] $\mu X.F=\mu X.(\Sigma_{i=1}^k\tau.(X+F''_i)+G)$
\item[$=$] $\mu X.(\tau.(X+\Sigma_{i=1}^kF''_i)+G)\hfill{\rm D4}$
\item[$=$] $\mu X.(\tau.\tau^*(\Sigma_{i=1}^kF''_i+G)+G)\hfill{\rm D3}$
\item[$=$]$\tau.\tau^*(\Sigma_{i=1}^kF''_i\{L/X\}+G\{L/X\})+G\{L/X\}\hfill{\rm R1}$
\end{enumerate}
where $L\equiv\mu X.(\tau.\tau^*(\Sigma_{i=1}^kF''_i+G)+G)$.
Now $L$ is a guarded expression
since $G$ and $F_1'',\ldots,F_k''$ are guarded expressions  and
$X$ is guarded in $G$ and $F_1'',\ldots,F_k''$.  Moreover since
$G$ is a standard guarded sum, so are
$G\{L/X\}$
and
$\tau.\tau^*(\Sigma_{i=1}^kF''_i\{L/X\}+G\{L/X\})+G\{L/X\}$.
\hfill\qed

\section{Quotient of Standard Equation System}

\begin{definition}\label{defofequationsystem}{\sl
A {\em recursive equation system} $S$ is a finite set of equations
$$\setof{X_i=F_i}{i=1,\ldots,n}
$$
where $X_1,\ldots,X_{n}\in\V$ are $n$ different variables, called the {\em formal variables} of $S$,
and $F_i\in\E$ for $i=1,\ldots,n$.

For $E\in\E$, $E$ is said to {\em provably solve (or satisfy)} the recursive equation system $S$
above for variable $X_k$ with $1\leq k\leq n$
if there are expressions $E_i$ for
$i=1,\ldots,n$ with $E$ being $E_{k}$, such that
$\ent E_i=F_i\{E_1/X_1,\ldots,E_n/X_n\}$  for
$i=1,\ldots,n$. 

Let $X,Y$ be two formal variables of a recursive equation system $S$,
$Y$ 
is said {\em $S$-unguarded} for $X$, written $X\,\tilde\triangleright_S Y$,
if  $Y$ occurs unguarded in the defining expression of
$X$ in $S$, i.e. $X=F_X\in S$ and $F_X\,\tilde\triangleright\,Y$.
A recursive equation system is said
{\em guarded} if $\tilde\triangleright_S$ is a well-founded
relation between the formal variables of $S$.
}
\end{definition}

\begin{theorem}\label{UniqueSolution} Let $S$ be a recursive equation system,
then for every formal variable of $S$
there is a provable solution. Moreover,
if $S$ is guarded, and both $D,E$ are provable solutions of $S$ for the same formal variable $X$, then $\ent E=D$,
i.e. every guarded recursive equation system has unique
solution up to provability.
\end{theorem}
{\bf Proof}. See \cite{milner89}, where only axioms valid in our inference system were used.
\hfill\qed

A particular kind of recursive equation system is standard equation system.

\begin{definition}{\sl
A {\em standard equation system} (or standard equation set), noted SES, is a guarded recursive equation system
in which
the right hand side of each equation has either of
the following forms
$$\begin{array}{ll}
(1)&\Sigma_{i=1}^na_iX_i+\Sigma_{j=1}^mW_j\\
(2)&\tau^*(\Sigma_{i=1}^na_i.X_i+\Sigma_{j=1}^mW_j)
\end{array}$$
where the $X_i$'s are formal variables of $S$, and $W_j$'s are not formal variables of $S$.
}
\end{definition}

\begin{definition}{\sl
Let $X,X'\in{\cal V}$ be two
formal variables of a guarded recursive equation system $S$.

$X$ and $X'$ are said to
have equivalent solutions of $S$ if whenever $E,E'\in\E$  are provable solutions
of $S$ for $X$ and $X'$ respectively then $E\approx_b^\vartriangle E'$,
and in this case we write $X\approx_SX'$.

$S$ is said to have {\em common provable solution} for
$X$ and $X'$ if there exists $E\in\E$ which
provable solves $S$ for $X$ as well as for $X'$.

A formal variable $X$ of $S$ is called a
{\em bottom variable} if
whenever $X\,\tilde\triangleright_S Y$ for some formal variable $Y$ of $S$ then $X\not\approx_SY$.}
\end{definition}

The following lemma shows that when $S$ is guarded, then $\approx_S$ is (as the symbol suggests) an equivalence
relation between the formal variables of $S$.
\begin{lemma} Let $S$ be a guarded recursive equation system.
\begin{enumerate}
\item $\approx_S$ is an equivalence
relation between the formal variables of $S$;
\item for any formal variable $X$ of $S$, $[X]$ contains a bottom variable, where
$[X]$ is the $\approx_S$-equivalence class containing $X$.
\end{enumerate}
\end{lemma}
{\bf Proof}. We first prove 1). Since $S$ is guarded, Theorem \ref{UniqueSolution}
guarantees unique solution for every formal variable $X$, which implies
reflexivity of $\approx_S$.
Symmetry is obvious by the symmetric phrase in the definition.
Transitivity easily follows from the transitivity of $\approx^\vartriangle_b$.

We now turn to 2). By the guardedness of $S$ we know that $\tilde\triangleright_S$ is a
well-founded relation between formal variables of $S$.
We prove the lemma by well-founded induction on $\tilde\triangleright_S$.
If $X$ is a bottom element, then of course $[X]$ contains $X$ which is a bottom element.
If $X$ is not a bottom element, then
there is a formal variable $X'$ such that $X'\in[X]$ and $X\tilde\triangleright_SX'$.
By the induction hypothesis $[X']$ contains a bottom element, thus $[X]$ contains a bottom element
since  $[X]=[X']$.
\hfill\qed
The notion of bottom variable resembles the notion of bottom elements of branching bisimulation
equivalence class introduced in \cite{GV90}.

The main purpose of this section is to use the quotient construction to
prove Theorem \ref{quotienting}, which states that
the $\approx_S$-equality between formal variables of a standard equation system $S$
implies common solution for the formal variables of a related equation system.
We need some preparation to prove the theorem.
The next three lemmas state some important properties of bottom variables of SES.
\begin{lemma}\label{bottomvariable0}{\sl Let $S$ be an SES, $X=F_X\in S$. If
$X$ is a bottom variable and $F_X\arrow{\tau}X'$, then $X'\not\approx_S X$.
}
\end{lemma}
{\bf Proof.} In this case $X\,\tilde\triangleright_S X'$, and since $X$ is a bottom variable, $X'\not\approx_S X$.
\hfill\qed

\begin{lemma}\label{bottomvariable}{\sl Let $S$ be an SES, $X, Y$ be two formal variables
of $S$ with $X=F_X,Y=F_Y\in S$. If
$X$ is a bottom variable and $X\approx_SY$, then
\begin{enumerate}
\item whenever $F_Y\vartriangleright W$ then $F_X\vartriangleright W$;
\item whenever $F_Y\arrow{a}Y'$ where either  $Y'\not\approx_S X$ or
$a\not=\tau$, then there exists $X'$ such that
$F_X\arrow{a}X'$ and $X'\approx_S Y'$.
\end{enumerate}
}
\end{lemma}
{\bf Proof.}  Let $X_1,\ldots,X_n$ be the formal variables of $S$,\\
$S=\{X_1=F_{X_1},\ldots,X_n=F_{X_n}\}$,
$D_{X_1},\ldots,D_{X_n}$ be a set of provable solutions of $S$ for $X_1,\ldots,X_n$ respectively,
i.e. $\ent D_{X_i}=F_{X_i}\setof{D_{X_i}/X_i}{i=1,\ldots,n}$ holds for $i=1,\ldots,n$
(according to Theorem \ref{UniqueSolution} there exist such solutions). Without loss
of generality let $X, Y$ be $X_1$ and $X_2$ respectively, thus $X_1$ is a bottom
variable. In particular, $D_{X_1}=^\vartriangle_bF_{X_1}\setof{D_{X_i}/X_i}{i=1,\ldots,n}$
and \\
$D_{X_2}=^\vartriangle_bF_{X_2}\setof{D_{X_i}/X_i}{i=1,\ldots,n}$
follow from the soundness of the inference system, and the condition
$X_1\approx_SX_2$ forces $D_{X_1}\approx^\vartriangle_bD_{X_2}$ to hold, thus
$$F_{X_1}\setof{D_{X_i}/X_i}{i=1,\ldots,n}\approx^\vartriangle_bF_{X_2}\setof{D_{X_i}/X_i}{i=1,\ldots,n}.$$

First we prove the following basic fact: whenever \\
$F_{X_1}\setof{D_{X_i}/X_i}{i=1,\ldots,n}\dobarrow{}G$ with $G\approx^\vartriangle_bD_{X_1}$, then\\
$G\equiv F_{X_1}\setof{D_{X_i}/X_i}{i=1,\ldots,n}$. We can prove this by induction on the length
of the $\arrow{\tau}$ transition sequence from $F_{X_1}\setof{D_{X_i}/X_i}{i=1,\ldots,n}$ to $G$.
If the length is zero, then $G$ does not make a move from $F_{X_1}\setof{D_{X_i}/X_i}{i=1,\dots,n}$,
clearly in this case $G\equiv F_{X_1}\setof{D_{X_i}/X_i}{i=1,\dots,n}$ holds. \\
If the length
is not zero, then there is $G'$ such that $F_{X_1}\setof{D_{X_i}/X_i}{i=1,\dots,n}\arrow{\tau}G'$,
$G'\dobarrow{}G$ and the number of transitions from
$G'$ to $G$ is one less than that from\\
$F_{X_1}\setof{D_{X_i}/X_i}{i=1,\dots,n}$, and since $D_{X_1}$ provably solves $S$,
$D_{X_1}=^\vartriangle_bF_{X_1}\setof{D_{X_i}/X_i}{i=1,\dots,n}$ follows from the
soundness of the inference system, thus \\
$F_{X_1}\setof{D_{X_i}/X_i}{i=1,\dots,n}
\approx^\vartriangle_bD_{X_1}\approx^\vartriangle_bG$,  we know from stutter lemma
that $G'\approx^\vartriangle_bG\approx^\vartriangle_b D_1$.
On the other hand, the form of the expression $F_{X_1}$ determines that there are two possible
ways for $F_{X_1}\{D_1/X_1,\dots,D_n/X_n\}$ to make a $\arrow{\tau}$ transition to $G'$:
either for some $k$ such that $F_{X_1}\arrow{\tau}X_k$ and $G'$ is $D_k$, or $F_{X_1}$ is a loop expression and
$G'\equiv F_{X_1}\setof{D_{X_i}/X_i}{i=1,\dots,n}$. Since $X_1$ is a bottom variable, $X_1\,\tilde\triangleright X_k$
would force $X_k\not\approx_SX_1$, and then $G'\equiv D_k\not\approx^\vartriangle_bD_1$, so
the first
choice is not possible. Then it must be the second choice, and in this case
$G'$ is $F_{X_1}\setof{D_{X_i}/X_i}{i=1,\dots,n}$, and there is a shorter sequence of $\arrow{\tau}$ transitions
to $G$, by the induction hypothesis in this case $G\equiv F_{X_1}\setof{D_{X_i}/X_i}{i=1,\dots,n}$.

With this basic fact, we proceed to prove 2) as follows. 1) can be proved in the
same way.
Suppose
$F_{X_2}\arrow{a}X_k$
where $a\not=\tau$ or $X_k\not\approx_SX_2$. Thus $a=\tau$ or $D_{X_k}\not\approx^\vartriangle_bD_{X_2}$, and
$F_{X_2}\setof{D_{X_i}/X_i}{i=1\ldots,n}\arrow{a}D_{X_k}$. Then
there must exist $G,G'$ such that $F_{X_1}\setof{D_{X_i}/X_i}{i=1\ldots,n}\dobarrow{}G$,
$G\arrow{a}G'$ such that $G\approx^\vartriangle_bD_{X_1}$ and $G'\approx^\vartriangle_bD_{X_i}$.
By the property proved above, it must be that $G\equiv F_{X_1}\setof{D_{X_i}/X_i}{i=1,\ldots,n}$, thus
$F_{X_1}\setof{D_{X_i}/X_i}{i=1,\ldots,n}\arrow{a}G'$,
and the form of the expression $F_{X_1}$ determines that there are two possible
ways for $F_{X_1}\{D_1/X_1,\dots,D_n/X_n\}$ to make a $\arrow{a}$ transition to $G'$:
either for some $l$ such that $F_{X_1}\arrow{a}X_l$ and $G'$ is $D_{X_l}$, or $a=\tau$ and
$F_{X_1}$ is a loop expression and
$G'$ is \\
$F_{X_1}\setof{D_{X_i}/X_i}{i=1,\dots,n}$. The second alternative is not possible, since that
would imply $D_{X_k}\approx^\vartriangle_bG'\approx^\vartriangle_bD_{X_2}$ and
$a=\tau$, contradicts the condition that $a\not=\tau$ or $X_k\not\approx_SX_2$.
Thus we found $X_l$ such that $F_{X_1}\arrow{a}X_l$, and since $D_{X_l}$ provably solves
$X_l$ and $D_{X_l}\approx^\vartriangle_bD_{X_k}$, hence $X_l\approx_SX_k$.\hfill\qed

\begin{lemma}\label{loopexp} Let $S$ be an SES, $X$ be a bottom variable with $X=F_X\in S$.
If $F_X$ is not a loop expression, then whenever $Y=F_Y\in S$ with $X\approx_SY$,
$F_Y$ is not a loop expression.
\end{lemma}
{\bf Proof.} Any solution to a bottom variable $X$ with $X=F_X\in S$ where $F_X$
is not a loop expression must not be divergent, while any solution to a variable $Y$ (bottom or not)
with $Y=F_Y\in S$ where $F_Y$ is a loop expression must be divergent, thus in this case
$X\not\approx_SY$.
\hfill\qed

\begin{definition}{\sl Let $S$ be an SES with the set of formal variables $V$, $X=F_X$ be an equation in $S$.
Define $F^0_X$ and $F^1_X$, called
the {\em stuttering derivative} and {\em non stuttering derivative} of $X$ in $S$ respectively, as the follows\\
$
F^0_X\!=\!\Sigma\setof{\tau.Y}{F_X\arrow{\tau}Y,X\approx_SY},$\\
$F^1_X\!=\!\Sigma\setof{a.Y}{F_X\!\arrow{a}\!Y,a\!\not=\!\tau\mbox{ or }X\!\not\approx_S\!Y}
+\Sigma\setof{W}{F_X\vartriangleright W}.
$}
\end{definition}

Intuitively, $F^0_X$ is a sum which consists of the $\tau$-derivatives of $F_X$ which do not change
equivalence class, while $F^1_X$ is a sum which consists of the rest of the derivatives plus the
free variables.
Here are some simple properties of derivatives.

\begin{proposition}\label{propder}
Let $S$ be an SES with formal variables in $V$,
$X=F_X, X'=F_{X'}$ be equations in $S$,
$\setof{E_Y/Y}{Y\in V}$ be a $\approx_S$-respecting
substitution on $V$ s.t. whenever $Y\approx_SY'$ then $E_Y\equiv E_{Y'}$.
\begin{enumerate}
\item if $F_X$ is not a loop expression, then\\
$\ent F_X\setof{E_Y/Y}{Y\in V}\\
=F^0_X\setof{E_Y/Y}{Y\in V}+F^1_X\setof{E_Y/Y}{Y\in V}$;
\item if $F_X$ is a loop expression, then\\
$\ent F_X\setof{E_Y/Y}{Y\in V}\\
=\tau^*(F^0_X\setof{E_Y/Y}{Y\in V}+F^1_X\setof{E_Y/Y}{Y\in V})$;
\item
$\ent F_X\setof{E_Y/Y}{Y\in V}\\
=F_X\setof{E_Y/Y}{Y\in V}+F^1_X\setof{E_Y/Y}{Y\in V}$;
\item if $X$ is a bottom variable then \\
$\ent F^0_X\setof{E_Y/Y}{Y\in V}={\bf 0}$, otherwise\\
$\ent F^0_X\setof{E_Y/Y}{Y\in V}=\tau.E_X$;
\item if $X\approx_SX'$, and $X$ is a bottom variable then \\
$\ent F^1_X\setof{E_Y/Y}{Y\in V}\\
=F^1_X\setof{E_Y/Y}{Y\in V}+F^1_{X'}\setof{E_Y/Y}{Y\in V}$.
\end{enumerate}
\end{proposition}
{\bf Proof.} To prove 1), note that when $F_X$ is not a loop expression
then $F^0_X+F^1_X$ and $F_X$ are two sums with the same set summands.
Thus
$F^0_X\setof{E_Y/Y}{Y\in V}+F^1_X\setof{E_Y/Y}{Y\in V}$ and
$F_X\setof{E_Y/Y}{Y\in V}$
are also two sums with the same set of summands, so by {\rm S1-S4}\\
$\ent F_X\setof{E_Y/Y}{Y\in V}\\
=F^0_X\setof{E_Y/Y}{Y\in V}+F^1_X\setof{E_Y/Y}{Y\in V}$.

2) can be proved in the same way.

3) easily follows from 1) and 2) by using ${\rm S3}$ and ${\rm D2}$.

To see 4), first note that by Lemma \ref{bottomvariable0}
$F^0_X$ is an empty sum when $X$ is a bottom variable. Thus
$F^0_X\setof{E_Y/Y}{Y\in V}$ is also an empty sum, and $\ent F^0_X\setof{E_Y/Y}{Y\in V}={\bf 0}$.
When $X$ is not a bottom element, then there exists $X'$ such that $X\approx_SX'$ and $F_X\arrow{\tau}X'$.
Then by the $\approx_S$-respecting property of the substitution, for all such $X'$ it holds that
$E_{X'}\equiv E_X$, hence $\ent F^0_X\setof{E_Y/Y}{Y\in V}=\tau.E_X$.

To prove 5), note that, according to Lemma \ref{bottomvariable}
when $X\approx_SX'$ and $X$ is a bottom variable, with the $\approx_S$-respecting
condition of the substitution,
any summand of $F^1_{X'}\setof{E_Y/Y}{Y\in V}$
is a summand of $F^1_X\setof{E_Y/Y}{Y\in V}$.
Thus by ${\rm S3}$\\
$\ent F^1_X\setof{E_Y/Y}{Y\in V}\\
=F^1_X\setof{E_Y/Y}{Y\in V}+F^1_{X'}\setof{E_Y/Y}{Y\in V}$.
\hfill\qed

\begin{definition}{\sl For a recursive equation set $S$, define a new recursive equation set
$\tau(S)=\setof{X=\tau.F_X}{X=F_X\in S}$.}
\end{definition}

Then $\tau(S)$ has the same set of
formal variables as $S$.

\begin{lemma}\label{abouttau00} Let $S$ be a recursive equation set, $X,Y$ be formal variables of $S$. Then
$X\,\tilde\triangleright_SY$ if and only if
$X\,\tilde\triangleright_{\tau(S)}Y$.
\end{lemma}
{\bf Proof.} Obvious.\hfill\qed

\begin{lemma}\label{abouttau} Let $S$ be an SES. Then
$\tau(S)$ is guarded.
\end{lemma}
{\bf Proof.} Note that an SES is a guarded recursive equation set, then the claim immediately
follows from Lemma \ref{abouttau00}. \hfill\qed

\begin{proposition}\label{Prequotienting} Let $S$ be an SES, $X$ be one of the formal variables.
If $E\in\E$ is a provable solution of $S$ for $X$
then $\tau.E$ is a provable solution of $\tau(S)$ for $X$.
\end{proposition}
{\bf Proof.} It easily follows from ${\rm T1}$ (Theorem \ref{T1asDerivedRule}).
\hfill\qed

We need the following lemma about successive substitution applied on a simple sum $F$
(Definition \ref{defofsimplesum}).
The lemma also holds for general expression $F$, which is the well-known
substitution lemma (for example Lemma 2.1 in \cite{Winskel91}). The simplified version is sufficient
for our purpose, and is easy to establish.
\begin{lemma}\label{substitutionlemmasimple} Let $F$ be a simple sum, $X_1,\ldots,X_n$ be $n$
variables, $E_1,\ldots,E_n$ be $n$ expressions, $Z_1,\ldots,Z_m$ be $m$ variables which do
not occur in $F$, $N_1,\ldots,N_m$ be $m$ expressions. Then \\
$F\{E_1/X_1,\ldots,E_n/X_n\}\{N_1/Z_1,\ldots,N_m/Z_m\}\equiv\\
F\{E_1\{N_1/Z_1,\ldots,N_m/Z_m\}/X_1,\\
\mbox{}\ldots,
E_n\{N_1/Z_1,\ldots,N_m/Z_m\}/X_n\}
$
\end{lemma}
{\bf Proof.} Can be proved by an easy induction on the number of summands in $F$.
\hfill\qed


We now arrive at the main result of this paper, we name it quotioning theorem because the
proof uses the quotient construction of an SES.
\begin{theorem}\label{quotienting} (Quotienting) Let $S$ be an SES, $X,X'$ be two formal variables of $S$.
If $X\approx_SX'$, then $\tau(S)$ has common provable solution for $X$ and $X'$.
\end{theorem}
{\bf Proof.}
Let $V$ be
the set of formal variables of $S$.
Since $S$ is an SES  which is a
guarded recursive equation system, $\approx_S$ is an equivalence relation on $V$,
we can assume that $\approx_S$ partitions $V$ into $n$ equivalence classes
$C_1,\ldots,C_n$. According to Lemma \ref{bottomvariable}, there exists a bottom variable in
each equivalence class, thus we can assume that $X_1,\ldots,X_n$ are
$n$ designated
bottom variables such that  $X_i\in C_i$ with
$X_i=F_{X_i}\in S$ for \\
$i=1,\ldots,n$.
For $X\in V$, we define the {\em index of} $X$, written $\iota(X)$,
such that $\iota(X)=i$
if $X\in C_i$.
Let $Z_1,\ldots,Z_n$
be $n$ variables which are not variables occurring in
any equation in $S$. We construct the following $\approx_S$-quotient equation system
of $S$:
$$S/\!\!\approx_S\,=\setof{Z_i={G_i}}{i=1,\ldots,n}
$$
where ${G_i}\equiv F_{X_i}\setof{Z_{\iota(Y)}/Y}{Y\in V}$. Then $S/\!\!\approx_S$
is a recursive equation system, and according to Theorem
\ref{UniqueSolution},
there exist $n$ expressions $B_{1},\ldots,B_{n}$
such that for $i=1,\ldots,n$ we have the following equality:
$$\ent B_{i}={G_{i}}\{B_{1}/Z_{1},\ldots,B_{n}/Z_{n}\}.\ \ \ \ \ (1)
$$
With this equality, we have that for each $i\in\{1,\ldots,n\}$:
\begin{itemize}
\item[$\ent$]$\!B_i=G_{i}\{B_{1}/Z_{1},\ldots,B_{n}/Z_{n}\}\hfill(1)$
\item[$=$]$\!F_{X_i}\setof{Z_{\iota(Y)}/Y}{Y\!\in\!V}\{B_{1}/Z_{1},\ldots,B_{n}/Z_{n}\}$\hfill def.of $G_{i}$
\item[$=$]$\!F_{X_i}\setof{Z_{\iota(Y)}\{B_{1}/Z_{1},\ldots,B_{n}/Z_{n}\}/Y}{Y\!\in\!V}$ Lemma \ref{substitutionlemmasimple}
\item[$=$]$\!F_{X_i}\setof{B_{\iota(Y)}/Y}{Y\!\in\!V}$
\end{itemize}
Thus for $i=1,\ldots,n$ we proved the following equality which will be used later
$$\ent B_i=F_{X_i}\setof{B_{\iota(Y)}/Y}{Y\!\in\!V}.\ \ \  (2)
$$

Our next step is to prove that for each $X\in V$ with \\
$X\!=\!F_X\in S$
it holds that
$$\ent \tau.F_X\setof{B_{\iota(Y)}/Y}{Y\in V}=\tau.F_{X_i}\setof{B_{\iota(Y)}/Y}{Y\in V},\ \ (3)$$
where $i=\iota(X)$, i.e.
$X_i$ is the designated bottom variable in $[X]$ and
$X_i=F_{X_i}\in S$.
To prove this we discuss four cases according to whether $X$ is a bottom variable and whether
$F_X$ is a loop expression by using Proposition \ref{propder} (note that $\setof{B_{\iota(Y)}/Y}{Y\in V}$ is clearly a
$\approx_S$-respecting substitution).

If $X$ is a bottom variable and $F_X$ is not a loop expression,
then by Lemma \ref{loopexp} $F_{X_i}$ is not a loop expression. So\\
$\ent \tau.F_X\setof{B_{\iota(Y)}/Y}{Y\in V}\\
=\tau.(F^0_X\setof{B_{\iota(Y)}/Y}{Y\in V}+F^1_X\setof{B_{\iota(Y)}/Y}{Y\in V})
$\\
\mbox{}\ \hfill 1) of proposition \ref{propder}\\
$=\tau.F^1_X\setof{B_{\iota(Y)}/Y}{Y\in V}$ \hfill 4) of prop. \ref{propder} and {\rm S4}\\
$=\tau.(F^1_X\setof{B_{\iota(Y)}/Y}{Y\in V}+
F^1_{X_i}\setof{B_{\iota(Y)}/Y}{Y\in V})$\\
\mbox{}\ \hfill 5) of proposition \ref{propder}, $X$ is a bottom variable\\
$=\tau.F^1_{X_i}\setof{B_{\iota(Y)}/Y}{Y\!\in\!V}$
\hfill 5) of prop. \ref{propder}, $X_i$ a bot. var.\\
$=\tau.(F^0_{X_i}\setof{B_{\iota(Y)}/Y}{Y\in V}+F^1_{X_i}\setof{B_{\iota(Y)}/Y}{Y\in V})
$\\
\mbox{}\ \hfill 4) of prop. \ref{propder} and {\rm S4}\\
$=\tau.F_{X_i}\setof{B_{\iota(Y)}/Y}{Y\in V}$
 \hfill 1) of proposition \ref{propder}

If $X$ is a bottom variable and $F_X$ is a loop expression,
then $F_{X_i}$ is also a loop expression (otherwise, since
$X_i$ is a bottom variable, by Lemma \ref{loopexp} $F_X$ is not a loop expression).
Then\\
$\ent \tau.F_X\setof{B_{\iota(Y)}/Y}{Y\in V}\\
=\tau.\tau^*(F^0_X\setof{B_{\iota(Y)}/Y}{Y\in V}+F^1_X\setof{B_{\iota(Y)}/Y}{Y\in V})
$\\
\mbox{}\ \hfill 2) of proposition \ref{propder}\\
$=\tau.\tau^*F^1_X\setof{B_{\iota(Y)}/Y}{Y\in V}$ \hfill 4) of prop. \ref{propder} and {\rm S4}\\
$=\tau.\tau^*(F^1_X\setof{B_{\iota(Y)}/Y}{Y\in V}+
F^1_{X_i}\setof{B_{\iota(Y)}/Y}{Y\in V})$\\
\mbox{}\ \hfill 5) of proposition \ref{propder}, $X$ is a bottom variable\\
$=\tau.\tau^*F^1_{X_i}\setof{B_{\iota(Y)}/Y}{Y\!\in\!V}$
\hfill 5) of prop. \ref{propder}, $X_i$ bot. var.\\
$=\tau.\tau^*(F^0_{X_i}\setof{B_{\iota(Y)}/Y}{Y\in V}+F^1_{X_i}\setof{B_{\iota(Y)}/Y}{Y\in V})
$\\
\mbox{}\ \hfill 4) of prop. \ref{propder} and {\rm S4}\\
$=\tau.F_{X_i}\setof{B_{\iota(Y)}/Y}{Y\in V}$
 \hfill 2) of proposition \ref{propder}

If $X$ is not a bottom variable and $F_X$ is a loop expression,
then $F_{X_i}$ is also a loop expression (otherwise, since
$X_i$ is a bottom variable, $F_X$ cannot be a loop expression).
Now\\
$\ent \tau.F_X\setof{B_{\iota(Y)}/Y}{Y\in V}\\
=\tau.\tau^*(F^0_X\setof{B_{\iota(Y)}/Y}{Y\in V}+F^1_X\setof{B_{\iota(Y)}/Y}{Y\in V})
$\\
\mbox{}\ \hfill 2) of proposition \ref{propder}\\
$=\tau.\tau^*(\tau.B_{\iota(X)}+F^1_X\setof{B_{\iota(Y)}/Y}{Y\in V})$\hfill 4) of prop. \ref{propder}\\
$=\tau.\tau^*(\tau.B_{i}+F^1_X\setof{B_{\iota(Y)}/Y}{Y\in V})$\hfill identity\\
$=\tau.\tau^*(\tau.F_{X_i}\setof{B_{\iota(Y)}/Y}{Y\!\in\!V}\!+\!F^1_X\setof{B_{\iota(Y)}/Y}{Y\!\in\!V})$
(2)
$=\tau.\tau^*(\tau.\tau^*(F^0_{X_i}\setof{B_{\iota(Y)}/Y}{Y\!\in\!V}+F^1_{X_i}\setof{B_{\iota(Y)}/Y}{Y\!\in\!V})\\
\mbox{}\ \ \ +F^1_X\setof{B_{\iota(Y)}/Y}{Y\!\in\!V})$\hfill 2) of proposition \ref{propder}\\
$=\tau.\tau^*(\tau.\tau^*F^1_{X_i}\setof{B_{\iota(Y)}/Y}{Y\!\in\!V}\\
\mbox{}\ \ \ +F^1_X\setof{B_{\iota(Y)}/Y}{Y\!\in\!V})$\hfill 4) of proposition \ref{propder}\\
$=\tau.\tau^*(\tau.\tau^*(F^1_{X_i}\setof{B_{\iota(Y)}/Y}{Y\!\in\!V}+F^1_{X}\setof{B_{\iota(Y)}/Y}{Y\!\in\!V})\\
\mbox{}\ \ \ +F^1_X\setof{B_{\iota(Y)}/Y}{Y\!\in\!V})$\hfill 5) of proposition \ref{propder}\\
$=\tau.(\tau.\tau^*(F^1_{X_i}\setof{B_{\iota(Y)}/Y}{Y\!\in\!V}+F^1_{X}\setof{B_{\iota(Y)}/Y}{Y\!\in\!V})\\
\mbox{}\ \ \ +F^1_X\setof{B_{\iota(Y)}/Y}{Y\!\in\!V})$\hfill {\rm D5}\\
$=\tau.(\tau.(\tau^*(F^1_{X_i}\setof{B_{\iota(Y)}/Y}{Y\!\in\!V}+F^1_{X}\setof{B_{\iota(Y)}/Y}{Y\!\in\!V})\\
\mbox{}\ \ \ +F^1_X\setof{B_{\iota(Y)}/Y}{Y\!\in\!V})+F^1_X\setof{B_{\iota(Y)}/Y}{Y\!\in\!V})
$\hfill {\rm D2}\\
$=\tau.(\tau^*(F^1_{X_i}\setof{B_{\iota(Y)}/Y}{Y\!\in\!V}+F^1_{X}\setof{B_{\iota(Y)}/Y}{Y\!\in\!V})\\
\mbox{}\ \ \ +F^1_X\setof{B_{\iota(Y)}/Y}{Y\!\in\!V})
$\hfill {\rm B}\\
$=\tau.\tau^*(F^1_{X_i}\setof{B_{\iota(Y)}/Y}{Y\!\in\!V}+F^1_{X}\setof{B_{\iota(Y)}/Y}{Y\!\in\!V})
$\hfill {\rm D2}\\
$=\tau.\tau^*F^1_{X_i}\setof{B_{\iota(Y)}/Y}{Y\!\in\!V}
$\hfill 5) of proposition \ref{propder}\\
$=\tau.\tau^*(F^0_{X_i}\setof{B_{\iota(Y)}/Y}{Y\!\in\!V}+F^1_{X_i}\setof{B_{\iota(Y)}/Y}{Y\!\in\!V})
$\\
$=\tau.F_{X_i}\setof{B_{\iota(Y)}/Y}{Y\in V}$
 \hfill 2) of proposition \ref{propder}

If $X$ is not a bottom variable and $F_X$ is not a loop expression,
then\\
$\ent \tau.F_X\setof{B_{\iota(Y)}/Y}{Y\in V}\\
=\tau.(F^0_X\setof{B_{\iota(Y)}/Y}{Y\in V}+F^1_X\setof{B_{\iota(Y)}/Y}{Y\in V})$\\
\mbox{} \hfill 1) of proposition \ref{propder}\\
$=\tau.(\tau.B_{\iota(X)}+F^1_X\setof{B_{\iota(Y)}/Y}{Y\in V})$\hfill 4) of prop. \ref{propder}\\
$=\tau.(\tau.B_{i}+F^1_X\setof{B_{\iota(Y)}/Y}{Y\in V})$\hfill identity \\
$=\tau.(\tau.F_{X_i}\setof{B_{\iota(Y)}/Y}{Y\!\in\!V}\!+\!F^1_X\setof{B_{\iota(Y)}/Y}{Y\!\in\!V})$\hfill $(2)$ \\
$=\tau.(\tau.(F_{X_i}\setof{B_{\iota(Y)}/Y}{Y\!\in\!V}+F^1_{X_i}\setof{B_{\iota(Y)}/Y}{Y\!\in\!V})\\
\mbox{}\ \ \ +F^1_X\setof{B_{\iota(Y)}/Y}{Y\!\in\!V})$\hfill  3) of proposition \ref{propder}\\
$=\tau.(\tau.(F_{X_i}\setof{B_{\iota(Y)}/Y}{Y\!\in\!V}+F^1_{X_i}\setof{B_{\iota(Y)}/Y}{Y\!\in\!V}\\
\mbox{}\ \ \ +F^1_X\setof{B_{\iota(Y)}/Y}{Y\!\in\!V})+F^1_X\setof{B_{\iota(Y)}/Y}{Y\!\in\!V})$\\
\mbox{}\ \hfill  5) of proposition. \ref{propder}\\
$=\tau.(F_{X_i}\setof{B_{\iota(Y)}/Y}{Y\!\in\!V}\!+\!F^1_{X_i}\setof{B_{\iota(Y)}/Y}{Y\!\in\!V})\\
\mbox{}\ \ \ +F^1_X\setof{B_{\iota(Y)}/Y}{Y\!\in\!V})$\hfill ${\rm B}$\\
$=\tau.F_{X_i}\setof{B_{\iota(Y)}/Y}{Y\!\in\!V}$\hfill 3) and 5) of proposition \ref{propder}\\
With this, we finished the discussion of all four cases.

To conclude the proof,
for each $X=F_X\in S$ with $\iota(X)=i$ we have\\
$\ent \tau.B_{\iota(X)}=\tau.B_i$\hfill identity\\
$=\tau.F_{X_i}\setof{B_{\iota(Y)}/Y}{Y\in V}$\hfill (2)\\
$=\tau.F_X\setof{B_{\iota(Y)}/Y}{Y\in V}$\hfill (3)\\
$=\tau.F_X\setof{\tau.B_{\iota(Y)}/Y}{Y\in V}.$\hfill ${\rm T1}$ (Theorem \ref{T1asDerivedRule})\\
So $\tau.B_{\iota(X)}$ is a provable solution to $\tau(S)$ for $X$. If $X\approx_SX'$, then
$\tau.B_{\iota(X')}$  is a provable solution to $\tau(S)$ for $X'$, but in this case
$\iota(X)=\iota(X')$, hence $\tau.B_{\iota(X)}\equiv\tau.B_{\iota(X')}$,
so $\tau(S)$ has  common provable solution for $X$ and $X'$. \hfill\qed

\section{The Completeness Proof}
Now we are prepared to prove the completeness of the axiomatisation.

\begin{theorem}\label{equationsystem}
Let $E\in\E$. If $E$ is guarded, then there is a standard equation system $S$ with a
formal variable $X$, such that
$E$ is a provable solution of $S$ for $X$.
\end{theorem}
{\bf Proof.}  It is proved by induction on the structure of $E$.

(i) $E\equiv\nil$. Take $S$ to be the single equation $X=\nil$.

(ii) $E\equiv W$. Take $S$ to be the single equation $X=W$.

(iii) $E\equiv a.E'$. By the induction hypothesis $E'$ provably solves a
standard equation set $S'$ for variable $X'$. Then add the equation $X=a.X'$
to $S'$ to form $S$, $E$ provably solves $S$ for $X$, and $S$ is a standard equation system.

(iv) $E\equiv E'+E''$. By the induction hypothesis $E'$ provably solves a
standard equation set $S'$ for $X'$ with $X'=F'\in S'$,
and $E''$ provably solves a
standard equation set $S''$ for $X''$ with $X''=F''\in S''$
(assume that the formal variables of $S'$ are distinct from those of $S''$). There are four cases
to discuss according to the forms of $F'$ and $F''$. If both $F'$ and $F''$ are standard sums
then take $S'\cup S''$ and add $X=F'+F''$ to form $S$ (with $X$ distinct from the formal
and free
variables of $S'$ and $S''$). If $F'$ is a standard sum while $F''\equiv\tau^*G''$
for some standard sum $G''$, then take
$S'\cup S''$ and add the equation $X=F'+\tau.X''+G''$ to form $S$ (with $X$ distinct from the formal
and free
variables of $S'$ and $S''$). Likewise for the case where $F''$ is a standard sum while $F'$ is
in loop form. If $F'\equiv\tau^*G'$ and $F''\equiv\tau^*G''$ for standard sums $G'$ and $G''$,
then take $S'\cup S''$ and add $X=\tau.X'+G'+\tau.X''+G''$ to form $S$
(with $X$ distinct from the formal
and free
variables of $S'$ and $S''$). With ${\rm D1}$ it is easy to see that in all the cases
$S$ is a standard equation set and that $E$ provably solves $S$ for $X$.

(v) $E\equiv\mu W'.E'$, with $W'$ guarded in $E'$. By the induction hypothesis
$E'$ provably solves a
standard equation set $S'$ for variable $X'$ with $X'=H\in S'$. We discuss two cases
according to whether $H$ is a standard sum or not. If $H$ is a standard sum,
take $\setof{Y=G\{H/W'\}}{Y=G\in S'}$ to form $S$, and
if $H\equiv\tau^*H'$ where $H'$ is a standard sum, take
$\setof{Y=G\{\tau.X'+H'/W'\}}{Y=G\in S'}$ to form $S$.
Since $W'$ is not a
formal variable of $S'$, in both cases $S$ is a standard equation set.
Also $E$ provably solves $S$ for $X$ (in the second case
${\rm D1}$ is used to show this).

(vi) $E\equiv\tau^*E'$. By the induction hypothesis $E'$ provably solves a
standard equation set $S'$ for  $X'$ with $X'=F'\in S'$. If $F'$ is
a standard sum,
then add the equation $X=\tau^*F'$
to $S'$ to form $S$, $E$ provably solves $S$ for $X$, and $S$ is a standard equation system.
If $F'\equiv\tau^*F''$ where $F''$ is a standard sum, then add the equation $X=F'$ to form $S$.
According to ${\rm D6}$,
$E$ provably solves $S$, and $S$ is a standard equation set.
\hfill\qed

\begin{lemma}\label{promotion}(Promotion) Let $E,F\in\E$ be guarded expressions.
If $E\approx^\vartriangle_bF$,
then $\ent \tau.E=\tau.F$.
\end{lemma}
{\bf Proof.} According to Theorem \ref{equationsystem}, there exist
standard equation systems
$S_1$ with a formal variable $X$ and $S_2$ with a formal variable $Y$
(assume they have disjoint sets of formal variables) such that $E$ and $F$
provably solve $S_1$ for variable $X$ and $S_2$ for variable $Y$ respectively.
Then it is clear that $S_1\cup S_2$ is an SES
with formal variables $X$ and $Y$, and that
$E$ and $F$ provably solve $S_1\cup S_2$ for $X$ and $Y$ respectively.
Then according to Proposition \ref{Prequotienting},
$\tau.E$ and $\tau.F$ provably solve $\tau(S_1\cup S_2)$ for $X$ and $Y$
respectively. If $E\approx^\vartriangle_bF$,
then $X\approx_{S_1\cup S_2}Y$, and according to Theorem \ref{quotienting}
$X$ and $Y$ have common solution in $\tau(S_1\cup S_2)$, i.e. there is an expression
$B$ which provably solves $\tau(S_1\cup S_2)$ for $X$ as well as for $Y$.
Since $S_1\cup S_2$ is an SES which is guarded, by Lemma \ref{abouttau}
$\tau(S_1\cup S_2)$ is also guarded,
according to Theorem \ref{UniqueSolution}
it has unique solution.  Now both $\tau.E$ and $B$ provably solve $\tau(S_1\cup S_2)$
for $X$ so $\ent \tau.E=B$, and both $\tau.F$ and $B$ provably solve $\tau(S_1\cup S_2)$
for $Y$ so $\ent \tau.F=B$, hence $\ent \tau.E=\tau.F$.
\hfill\qed

\begin{theorem} Let $E,F\in\E$. If $E=^\vartriangle_bF$ then $\ent E=F$.
\end{theorem}
{\bf Proof.} First we show that in this case $\ent E+F=F$.
By Theorem \ref{guardedness} there exist guarded expressions
$E_1,\ldots,E_n,$
and variables $W_1,\ldots,W_m,$
such that
\begin{enumerate}
\item[$\ent$] $E=\Sigma_{i=1}^na_i.E_i+\Sigma_{j=1}^mW_j.\hfill(A)$
\end{enumerate}
Thus in order to prove $\ent E+F=F$ we only need to show that
$\ent\Sigma_{i=1}^na_i.E_i+\Sigma_{j=1}^mW_j+F=F,$
and we will do this by induction on $m+n$. If $m+n=0$,
by ${\rm S4}$ obviously this holds. If $m>0$, then
$\Sigma_{i=1}^na_i.E_i+\Sigma_{j=1}^mW_j\vartriangleright W_m$.
Since $\Sigma_{i=1}^na_i.E_i+\Sigma_{j=1}^mW_j=^\vartriangle_bE=^\vartriangle_bF$ by
$(A)$ and the soundness of the proof system and $E=^\vartriangle_bF$,
it follows that $F\vartriangleright W_m$. Thus $\ent F=F+W_m$ follows
from Lemma \ref{Summand}. Now\\
$\ent \Sigma_{i=1}^na_i.E_i+\Sigma_{j=1}^mW_j+F\\
=\Sigma_{i=1}^na_i.E_i+\Sigma_{j=1}^{m-1}W_j+F+W_m\hfill{\rm S1}\\
=\Sigma_{i=1}^na_i.E_i+\Sigma_{j=1}^{m-1}W_j+F\hfill\mbox{Lemma \ref{Summand}}\\
=F\hfill\mbox{ind. hyp.}$\\
If $n>0$, then
$\Sigma_{i=1}^na_i.E_i+\Sigma_{j=1}^mW_j\arrow{a_n}E_n$. By
$(A)$ and the soundness of the proof system,
$\Sigma_{i=1}^na_i.E_i+\Sigma_{j=1}^mW_j=^\vartriangle_bF$,
it follows that $F\arrow{a_n}F'$ with $F'\approx^\vartriangle_bE_n$. Thus $\ent F=F+a_n.F'$ follows
from Lemma \ref{Summand}. Now\\
$\ent \Sigma_{i=1}^na_i.E_i+\Sigma_{j=1}^mW_j+F\\
=\Sigma_{i=1}^{n-1}a_i.E_i+\Sigma_{j=1}^{m}W_j+F+a_n.E_n\hfill{\rm S1}\\
=\Sigma_{i=1}^{n-1}a_i.E_i+\Sigma_{j=1}^{m}W_j+F+a_n.\tau.E_n\hfill{\rm T1}\\
=\Sigma_{i=1}^{n-1}a_i.E_i+\Sigma_{j=1}^{m}W_j+F+a_n.\tau.F'\hfill\mbox{Lemma }\ref{promotion}\\
=\Sigma_{i=1}^{n-1}a_i.E_i+\Sigma_{j=1}^{m}W_j+F+a_n.F'\hfill{\rm T1}\\
=\Sigma_{i=1}^{n-1}a_i.E_i+\Sigma_{j=1}^{m}W_j+F\hfill\mbox{Lemma \ref{Summand}}\\
=F\hfill\mbox{ind. hyp.}$

In the same way we can show $\ent E+F=F$, hence \\
$\ent E=E+F=F$.
\hfill\qed

\section{Conclusion and Future Work}
In this paper we presented a complete axiomatisation for divergence-preserving branching
congruence of finite-state behaviours.
Also,
along the way
of proving soundness we
identified three techniques for establishing divergence-preserving bisimulation equivalence and congruence:
the ${\cal B}(\approx^\vartriangle_b)$ technique (proof of Lemma \ref{branchingaxiom}),
the progressing branching bisimulation technique (proof of Lemma \ref{progtechnique}),
and the strong bisimulation up to $\approx^\vartriangle_b$
technique (proof of Lemma \ref{soundnessofR2for=1}).
Since they help to relieve one off the burden of showing divergence preservation,
these techniques enrich the theory of
divergence-preserving branching bisimulation, and could be
useful in other works. In \cite{aceto96} Aceto et al.
studied complete axiomatisations
for (divergence-blind) weak congruence, delay congruence, and $\eta$-congruence besides
branching congruence. These other congruences also have corresponding divergence-preserving version
similar to divergence-preserving branching congruence. We hope that the result of this paper may
help to establish sound and complete axiomatisations for these divergence-preserving congruences.

\section*{Acknowledgment}
The authors would like to thank David N. Jansen for proof reading a draft
of the paper, and the anonymous referees for suggestions of improvement. The work has been supported by the
CAS-INRIA major project No. GJHZ1844, and by NSFC under grants No. 61836005 and No. 62072443.






%

\end{document}